\documentclass[final,5p,times,twocolumn,sort&compress,nopreprintline]{elsarticle}
\biboptions{square,sort&compress,comma,numbers}

\usepackage{multirow}

\usepackage[colorlinks]{hyperref}
\usepackage[table,xcdraw,dvipsnames]{xcolor}
\usepackage[normalem]{ulem}

\usepackage[latin1]{inputenc}
\usepackage[english]{babel}
\usepackage{amssymb,amsmath,amsthm,cancel,graphicx,xcolor}
\usepackage{booktabs}
\usepackage{mathrsfs}

\usepackage{color}

\newcommand{\gsim}{\gtrsim}

\let\originalleft\left
\let\originalright\right
\renewcommand{\left}{\mathopen{}\mathclose\bgroup\originalleft}
\renewcommand{\right}{\aftergroup\egroup\originalright}

\def\beq{\begin{equation}}  
\def\eeq{\end{equation}}
\def\({\left(}
\def\){\right)}
\def\[{\left[}
\def\]{\right]}

\def\eq#1{{Eq.~(\ref{#1})}}
\def\eqs#1#2{{Eqs.~(\ref{#1})-(\ref{#2})}}
\def\fig#1{{Fig.~\ref{#1}}}

\def\sect#1{{Sect.~\ref{#1}}}

\def\vev#1{\left\langle #1\right\rangle}

\def\Tr{\mbox{Tr}\,}
\def\tr{\mbox{Tr}\,}

\def\wt#1{\widetilde{#1}}

\renewcommand{\bar}{\overline}

\newcommand{\X}{{\cal X}}
\newcommand{\U}{{\rm U}}

\newcommand{\qL}{{q_L}}
\newcommand{\uR}{{u_R}}
\newcommand{\dR}{{d_R}}
\newcommand{\lL}{{\ell_L}}
\newcommand{\eR}{{e_R}}

\newcommand{\GeV}{\,\mathrm{GeV}}
\newcommand{\TeV}{\,\mathrm{TeV}}

\renewcommand{\b}{\beta}

\newcommand{\SU}{{\rm SU}}

\newcommand{\PQ}{{\rm PQ}}

\newcommand{\BSM}{{\rm BSM}}

\newcommand{\1}{{\textbf{1}}}

\definecolor{rosso}{cmyk}{0,1,1,0.4}
\definecolor{rossos}{cmyk}{0,1,1,0.55}
\definecolor{rossoc}{cmyk}{0,1,1,0.2}
\definecolor{blu}{cmyk}{1,1,0,0.3}
\definecolor{blus}{cmyk}{1,1,0,0.6}
\definecolor{bluc}{cmyk}{1,1,0,0.1}
\definecolor{verde}{cmyk}{0.92,0,0.59,0.25}
\definecolor{verdec}{cmyk}{0.92,0,0.59,0.15}
\definecolor{verdes}{cmyk}{0.92,0,0.59,0.4}

\graphicspath{{figs/}} 

\journal{arXiv}

\begin{document}

\begin{frontmatter}

\title{
Running effects on QCD axion phenomenology  
}


\author[label1,label2]{Luca Di Luzio} 
\author[label3,label4]{Maurizio Giannotti}
\author[label5]{Federico Mescia} 
\author[label6,label7]{Enrico Nardi} 
\author[label5]{Shohei Okawa} 
\author[label8,label9]{Gioacchino Piazza} 
\address[label1]{Istituto Nazionale di Fisica Nucleare (INFN), Sezione di Padova, Via F.~Marzolo 8, 35131 Padova, Italy}
\address[label2]{Dipartimento di Fisica e Astronomia `G.~Galilei', Universit\`a di Padova, Via F.~Marzolo 8, 35131 Padova, Italy}
\address[label3]{Department of Chemistry and Physics, Barry University, 11300 NE 2nd Ave., Miami Shores, FL 33161, USA}
\address[label4]{Centro de Astropart{\'i}culas y F{\'i}sica de Altas Energ{\'i}as (CAPA), Universidad de Zaragoza, Zaragoza, 50009, Spain}
\address[label5]{Departament de F\'isica Qu\`antica i Astrof\'isica, Institut de Ci\`encies del Cosmos (ICCUB), \\ 
Universitat de Barcelona, Mart\'i i Franqu\`es 1, E-08028 Barcelona, Spain}
\address[label6]{Laboratory of High Energy and Computational Physics, HEPC-NICPB,  R\"avala 10, 10143 Tallinn, Estonia }
\address[label7]{Istituto Nazionale di Fisica Nucleare, Laboratori Nazionali di Frascati, C.P.~13, 00044 Frascati, Italy}
\address[label8]{IJCLab, P\^{o}le Th\'{e}orie (B\^{a}t.~210), CNRS/IN2P3 et Universit\'{e}  Paris-Saclay, 91405 Orsay, France}
\address[label9]{Physik-Institut, Universit\"{a}t Z\"{u}rich, CH-8057 Z\"{u}rich, Switzerland}

\begin{abstract}
We study the impact of renormalization group effects on QCD axion phenomenology. 
Focusing on the DFSZ model, we argue that the relevance of running effects for the axion couplings crucially depends on the scale where the heavier Higgs scalars are integrated out. 
We study the impact of these effects on astrophysical and cosmological bounds as well as on the sensitivity of helioscopes experiments such as IAXO and XENONnT, showing that they can be sizable even in the most conservative case in which the two Higgs doublets remain as light as the TeV scale. 
We provide simple analytical expressions that accurately fit the numerical solutions of the renormalization group equations as a function of the mass scale of the heavy scalars. 
\end{abstract}

\end{frontmatter}

\thispagestyle{plain}

\tableofcontents


\section{Introduction}
\label{sec:intro}

Axions are an intrinsic prediction of the Peccei-Quinn (PQ) mechanism~\cite{Peccei:1977hh,Peccei:1977ur}, which remains, after over four decades, the most appealing solution to the strong CP problem. 
This problem arises because Quantum Chromodynamics (QCD) predicts CP violating effects 
that are not observed experimentally. 
The PQ mechanism involves a new global  chiral symmetry U(1)$_\PQ$, 
which is anomalous under QCD and spontaneously broken at a large energy scale $f_a$ 
(PQ scale). 
The axion is the Nambu-Goldstone boson associated with the spontanous breaking of this
symmetry~\cite{Weinberg:1977ma,Wilczek:1977pj}, and is characterised by the fact that 
all its interactions are inversely proportional to $f_a$.
Although the original Weinberg-Wilczek model~\cite{Weinberg:1977ma,Wilczek:1977pj},
in which the scale of PQ breaking coincides with the electroweak (EW) symmetry breaking scale, was quickly ruled out, 
new viable models emerged early on, in which the PQ scale can be 
arbitrarily high so that   all axion interactions can be sufficiently suppressed, 
 yielding the so-called {\it invisible axion}. 
Two examples are particularly appealing for their simplicity: 
the KSVZ or hadronic axion~\cite{Kim:1979if,Shifman:1979if} and 
the DFSZ axion~\cite{Zhitnitsky:1980tq,Dine:1981rt}. 
The main difference between KSVZ and DFSZ-type axions is that the former do not couple to ordinary quarks and leptons at the tree level. 
Though many other possible axion models have been considered in the literature (see Ref.~\cite{DiLuzio:2020wdo} for a comprehensive overview),
the two above-mentioned models are by far the most studied ones and are  universally regarded as benchmark QCD axion models.
In recent years continuous progress in experimental technologies 
has  brought within reach the possibility of detecting in terrestrial experiments 
the invisible axions arising in these models.
This has stimulated a tremendous interest in this field, with several new theoretical and phenomenological studies, as well as a wealth of new experimental proposals 
(see \cite{Irastorza:2018dyq,Sikivie:2020zpn} for recent reviews). 
In the meanwhile, ongoing experiments have already started to probe the 
benchmark KSVZ/DFSZ axion models, and in the coming decades they will dig deep into the relevant parameter space region. 
From the theory side, this calls for 
the development of ``precision axion physics'', 
which will turn out to be crucial in the case of an  
axion discovery. Indeed, from a given determination 
of the low-energy axion couplings to photons 
and other matter fields (such as electrons and nucleons)
one would like to infer the structure of 
the high-energy theory, 
that is the ultraviolet (UV) 
completion of the axion effective field theory (EFT). 
This step is highly non-trivial, since it 
entails a large separation of scales, 
from the typical low-energy scale of axion experiments 
up to the PQ scale, $f_a \gtrsim 10^{8}$ GeV. 
Hence, axion related physical quantities, as for example 
the axion couplings to Standard Model (SM) fermions,  
are potentially affected by large radiative corrections, 
which can induce large deviations from the tree-level expressions.
In the case of the KSVZ model, it was 
pointed out long ago \cite{Srednicki:1985xd,Chang:1993gm} that 
although the axion coupling to electrons is zero at 
tree level, a non-zero electron coupling can be sourced via loop 
corrections by the axion-photon coupling and, more recently, it was shown 
that the leading correction to this  coupling is generated at even higher orders  
via the anomalous axion coupling to gluons~\cite{Choi:2021kuy}.
Nowadays, the full one-loop anomalous dimensions for the $d=5$ axion effective 
Lagrangian have been computed~\cite{Choi:2017gpf,Chala:2020wvs,Bauer:2020jbp,Bonilla:2021ufe}, 
while running effects have been 
investigated for the benchmark 
DFSZ/KSVZ axion models
in Ref.~\cite{Choi:2021kuy},  
and for the so-called astrophobic axion models (which feature non-universal PQ charges~\cite{DiLuzio:2017ogq,Bjorkeroth:2019jtx})   in Ref.~\cite{DiLuzio:2022tyc}. 

The purpose of this work is to study QCD axion phenomenology 
in light of renormalization group (RG) effects, 
focussing for definiteness on the mass window
$m_a \in [\text{meV}, \text{eV}]$.
This region of  parameter space 
shows a remarkable complementarity among the existing bounds on the
different axion couplings, namely to photons, 
electrons, nucleons and pions, stemming from 
helioscope searches, as well as 
from astrophysics and cosmology. It is therefore an ideal 
playground where to investigate the consequences of running effects for QCD axion phenomenology. 
In particular, we will 
focus on the large corrections induced by the top Yukawa coupling, 
which apply to a large class of axion models where the SM fermions are charged 
under the $\U(1)_{\rm PQ}$ symmetry. 
A paradigmatic example is the universal DFSZ model, 
which features two Higgs doublets
and one SM singlet scalar, 
and whose axion parameter space  
at tree level depends solely on 
$m_a$ and $\tan\beta$. 
However, top-Yukawa radiative corrections induce a 
logarithmic dependence of the effective axion couplings 
on the mass scale of the heavy scalar degrees of freedom 
of the two Higgs doublet model (2HDM) (the issue of large logarithmic corrections is well known in the 2HDM literature, see e.g.~\cite{Branco:2011iw,Krause:2016oke})
that can range from about $1$ TeV up to the PQ scale, $f_a$.  
As we shall see, these corrections are often large and may 
skew the parameter space region that is effectively probed by 
terrestrial experiments and by astrophysical/cosmological observations. 

The paper is organized as follows. 
In \sect{sec:runningcoupl} we discuss the structure 
of top-Yukawa radiative corrections, 
and we provide approximate analytical expressions
for the dependence  of the axion couplings on these corrections. 
\sect{sec:runningpheno} is devoted to study the impact of running effects on 
QCD axion phenomenology, including the consequences for astrophysical and 
cosmological limits as well as for the sensitivity of future axion experiments. 
We conclude in \sect{sec:concl}. 
Details on the solutions to the RG equations are provided in \ref{sec:RGEfit}.

\section{Running QCD axion couplings}
\label{sec:runningcoupl}
 
Of central interest for axion phenomenology are 
the axion couplings to photons and matter fields (electrons, nucleons,   
as well as other hadrons relevant for axion production). They are defined 
by the following interaction Lagrangian: 
\begin{align}
\label{eq:interactions}
\mathcal{L}_a &= C_{\gamma} \frac{\alpha}{8\pi} \frac{a}{f_a} 
F^{\mu\nu} \tilde F_{\mu\nu} + 
\sum_{f = p,\,n,\,e} C_{f} \frac{\partial_\mu a}{2 f_a} \bar f \gamma^\mu \gamma_5 f \nonumber \\ 
&+ C_{\pi} \frac{\partial_\mu a}{f_a f_\pi} (2 \partial^\mu \pi^0 \pi^+ \pi^- - \pi^0 \partial^\mu \pi^+ \pi^-  - \pi^0 \pi^+ \partial^\mu \pi^-) \nonumber \\
&+ C_{\pi N} \frac{\partial_\mu a}{2f_af_\pi}(i\pi^+\Bar{p}\gamma^\mu n-i\pi^-\Bar{n}\gamma^\mu p) \nonumber \\ 
&+C_{N\Delta} \frac{\partial^\mu a}{2f_a} \left(\Bar{p}\,\Delta^+_\mu +\overline{\Delta^+_\mu}\,p+\Bar{n}\,\Delta^0_\mu+\overline{\Delta^0_\mu}\,n\right)
+ \ldots \, , 
\end{align}
where $F_{\mu\nu}$  denotes  
the electromagnetic field strength,   
$\tilde F_{\mu\nu}= \frac{1}{2} \epsilon_{\mu\nu\rho\sigma} F^{\rho\sigma}$ (with $\epsilon^{0123}=-1$) 
its dual, $f = p,n,e$ runs over low-energy matter fields, and $C_{\gamma,\,f,\,\pi,\, \pi N,\, N\Delta}$ are   $\mathcal{O}(1)$ dimensionless coefficients. 
The axion-pion coupling in the second line   of \eq{eq:interactions} 
(with $f_\pi = 92.1(8)$ MeV \cite{ParticleDataGroup:2020ssz} 
 the pion decay constant) is of phenomenological relevance 
for thermal axion production in the early Universe, 
while the axion contact interactions with pions and nucleons (third line) 
and with  $\Delta$-resonances (fourth line) are important 
for axion production in Supernovae (SNe).
The ellipses stand for other possible 
axion interaction terms which will not be considered in this paper.

In the context of axion phenomenology, 
one usually employs the dimensional couplings 
$g_{a\gamma} = \frac{\alpha}{2\pi} \, C_\gamma / f_a$ and 
$g_{af} = C_f \, m_f / f_a$. 
In particular, $C_\gamma = E/N - 1.92(4)$, where $E/N$ 
is the ratio between the electromagnetic and QCD anomalies 
of the PQ current (for typical values in 
concrete axion models, see e.g.~\cite{DiLuzio:2016sbl,DiLuzio:2017pfr}). 
Note that 
to a very good approximation 
the anomalous axion-photon coupling
is insensitive
to running effects, with first corrections appearing  at three loops \cite{Bauer:2020jbp}. 
Moreover, mass dependent 
corrections to the effective axion-photon 
coupling
are safely negligible for $m_a \ll m_e$ 
since they scale 
at most as 
$(m_a / m_e)^2$ -- see e.g. Ref.~\cite{Bauer:2017ris}.   
Hence, in the following, we will only 
focus on radiative corrections to the 
axion couplings to electrons and hadrons. 

Axion-hadron interactions 
can be expressed in terms of the model-independent 
axion gluon coupling (which fixes the absolute 
normalization in terms of $f_a$) and the axion couplings 
to quark fields, $q=u,d,s,c,b,t$, 
defined via the Lagrangian 
term\footnote{In this work we focus on universal 
axion models, so that the axial-vector 
currents in \eq{eq:axialvector} are flavor diagonal.
However, 
as long as the PQ charges of different generations 
are not hierarchical,  
most of the considerations related 
to top-Yukawa 
running effects, to be discussed below, 
apply as well  to non-universal axion models 
(see e.g.~Ref.~\cite{DiLuzio:2022tyc}).} 
\beq 
\label{eq:axialvector}
C_q \frac{\partial_\mu a}{2 f_a} 
\bar q \gamma^\mu \gamma_5 q \, . 
\eeq 
In terms of the latter, 
the axion couplings to 
hadrons defined in \eq{eq:interactions} read 
(see e.g.~\cite{diCortona:2015ldu,DiLuzio:2020wdo,Choi:2021ign,Ho:2022oaw})
\begin{align}
\label{eq:CpDelta}
C_p &=
C_u \Delta_u + C_d \Delta_d + C_s \Delta_s 
-\left(\frac{\Delta_u}{1+z} + \frac{z \, \Delta_d}{1+z} \right) \, ,
\\
\label{eq:CnDelta}
C_n &=
C_d \Delta_u + C_u \Delta_d + C_s \Delta_s 
-\left(\frac{z \, \Delta_u}{1+z} + \frac{\Delta_d}{1+z} \right) \, , \\
\label{eq:defCapi}
C_{\pi} &= - \frac{1}{3}\(C_u-C_d -\frac{1-z}{1+z}\) 
\, , \\
\label{eq:defCNDelta}
C_{\pi N} &= -\frac{3}{\sqrt{2}} C_{\pi} \, , 
\qquad 
C_{N\Delta} =\frac{3\sqrt{3}}{2} C_{\pi} \, g_A \,,
\end{align}
where $C_{u,\,d,\,s} = C_{u,\,d,\,s} (2\GeV)$ 
are low-energy couplings evaluated 
at the scale $\mu = 2\GeV$ 
by numerically solving the RG equations  
from the boundary values $C_{u,\,d,\,s}(f_a)$
(see below), 
$\Delta_{u,\,d,\,s}$ denote the nucleon matrix elements of 
the quark axial-vector currents. 
In particular,
$g_A \equiv \Delta_u - \Delta_d = 
1.2754(13)$ 
from $\beta$-decays \cite{ParticleDataGroup:2020ssz},  
$\Delta_u =  0.847(18)(32)$, $\Delta_d =-0.407(16)(18)$ and $\Delta_s = -0.035(6)(7)$ 
(at 2 GeV in the $\overline{\text{MS}}$ scheme)
are the 
$N_f=2+1$ FLAG average \cite{FLAG2023}, 
that is dominated by the results of \cite{Liang:2018pis}, 
and $z = m_u(2 \GeV)/m_d(2 \GeV) = 0.49(2)$ \cite{FlavourLatticeAveragingGroupFLAG:2021npn}.
Combining lattice values with the high-precision determination of $g_A$, we obtain the weighted averages $\Delta_u =  0.858(22)$, $\Delta_d =-0.418(22)$ and $\Delta_s = -0.035(9)$. 

Running effects on the low-energy  couplings of the axion 
to first generation SM fermions 
can be parametrized as\footnote{In 
universal axion models 
the same corrections apply to each generation.
} 
(see e.g.~\cite{Choi:2021kuy})
\begin{align}
\label{eq:CuCdCe}
C_u (2 \, \text{GeV}) &= C_u (f_a) + \Delta C_u \, , \\ 
C_d (2 \, \text{GeV}) &= C_d (f_a) + \Delta C_d \, , \\ 
C_e (m_e) &= C_e (f_a) + \Delta C_e \, , 
\end{align}
with 
\beq  
\label{eq:Delta_C_Psi}
\Delta C_\Psi \simeq r^t_\Psi (m_{\rm BSM})\; C_t (f_a) \, ,
\eeq
and $\Psi = u,d,e$. 
The parameter $r^t_\Psi (m_{\rm BSM})$ encodes the RG 
correction approximated by taking only the top-Yukawa contribution,
and depends logarithmically on the 
parameter $m_{\rm BSM} \simeq m_{H,\, A,\, {H^+}}$ 
that denotes collectively the mass scale of the heavy scalar degrees of freedom
(we implicitly assume for the heavy modes of the scalar doublets 
the decoupling limit~\cite{Gunion:2002zf}, 
in which all the heavy masses are approximately degenerate).
The $m_{\rm BSM}$ scale depends on the 
structure of the DFSZ scalar potential, 
whose details 
(see e.g.~\cite{Bertolini:2014aia,Espriu:2015mfa}) 
are not crucial for the calculation of the axion RG 
equations, 
and we take it to range from about $1$ TeV 
(the approximate lower bound 
as set by LHC searches for new heavy scalars) 
up to $f_a$.  

Note that as long as the couplings are considered at a 
renormalization scale $\mu$ above  $m_{\rm BSM}$ 
there are no top-Yukawa running effects. 
This is because 
in this regime the axion couplings to the SM fermions 
correspond to the global charges of the  PQ current, 
which is classically conserved, and thus they do not renormalize.
For   $\mu < m_{\rm BSM}$ we enter a different regime, 
in which Higgs doublets with different PQ charges mix 
to give rise to  heavy scalars (which are integrated out) and to 
the light Higgs, that has no well-defined charge. In this effective theory 
there is no more a conserved PQ current, and running effects for the axion-fermion couplings can kick in. This is the reason why the largest RG effects appear when the BSM scale is taken at the largest possible scale 
$ m_{\rm BSM} \sim f_a$. Contrary, when the 2HDM structure keeps holding all the way down to the TeV scale, running effects are much less sizeable.

In \ref{sec:RGEfit} we provide
a fit to $r^t_\Psi (m_{\rm BSM}$) obtained  
by interpolating the 
numerical solution to the RG equations 
(cf.~\eqs{eq:rt3fit}{eq:rt0fit} 
and Tab.~\ref{tab:rpsit}). 
Taking, for instance, $m_{\rm BSM} = f_a = 10^{10}$ GeV one finds
\begin{align}
\label{eq:Cu2}
C_u (2 \, \text{GeV}) &\simeq C_u (f_a) - 0.264 \, C_t (f_a) \, , \\ 
\label{eq:Cd2}
C_d (2 \, \text{GeV}) &\simeq C_d (f_a) + 0.266 \, C_t (f_a) \, , \\ 
\label{eq:Ceme}
C_e (m_e) &\simeq C_e (f_a) + 0.265 \, C_t (f_a) \, . 
\end{align}

\subsection{Analytical understanding of RG running effects}
\label{sec:runningunderstanding}

To understand the phenomenological impact of the RG corrections 
to the axion couplings, and to compare it with the current experimental sensitivity, 
it is convenient to provide some analytical approximations.
To this aim, it is advantageous to
introduce the iso-scalar ($C_0$) and iso-vector ($C_3$) {nuclear}  couplings (see also \cite{DiLuzio:2022tyc}), defined as follows: 
\begin{align} 
\label{eq:CppCn}
    C_0= \frac{1}{2} \left(C_p + C_n\right)  &=  
    \frac{1}{2} \left(\Delta_u+\Delta_d\right)  \left( C_u  + C_d  - 1\right) - 
    \Delta_s C_s \, , \\
\label{eq:CpmCn}
    C_3=  \frac{1}{2}  (C_p -C_n) &= 
    \frac{g_A}{2} \left(C_u  - C_d  - 
    \frac{1-z}{1+z}
    \right) \, ,
\end{align}
where the right-hand sides  are obtained from 
the expressions for $C_{p,\,n}$ 
given in \eqs{eq:CpDelta}{eq:CnDelta}. 
From \eqs{eq:defCapi}{eq:defCNDelta} 
we see 
that all the other couplings are proportional to the iso-vector combination   $C_3$:
$C_\pi = -\frac{2}{3}\, g_{A}^{-1}  C_3 $, 
$C_{\pi N}=\sqrt{2}\, g_{A}^{-1}\,C_{3}$, 
$C_{N\Delta}=-\sqrt{3}\,C_{3}$.

The RG correction to 
the iso-vector combination 
$\Delta C_3\simeq 0.64\, C_t(f_a)\, ( r_u^t-r_d^t )$ may be sizeable, 
with the exact value depending on the $m_{\rm  BSM}$ scale.
An excellent fit to the combination $r_u^t-r_d^t$, for $m_{\rm  BSM}$ in the range 1 TeV to $10^{18}$ GeV, is given by
\begin{align}
\label{eq:r_3}
r_u^t-r_d^t \approx -0.54 \ln \left( \sqrt{x} - 0.52 \right)\,,
\end{align}
with $x=\log_{10} \left( m_{\rm  BSM}/{\rm GeV} \right)$.
This expression reproduces our numerical results with a precision better than $2\%$ (see~\ref{sec:RGEfit}).
Then, in the relevant range for $m_{\rm  BSM}$, we have $0.3\lesssim |r_u^t-r_d^t| \lesssim 1$. 
Since in universal axion models we expect $C_3\sim C_t$  (the exact relation depending on the model parameters),  
we can conclude that $\Delta C_3/C_3$ can be of the order of a few~10\%, 
and even larger.
For example, in the case of the DFSZ axion (to be discussed below), we find 
\begin{align}
\label{eq:deltaC3}
\left|\frac{\Delta C_3}{C_3}\right|_{\rm DFSZ}\simeq \frac{0.5}{\tan\beta} \,\left( r_u^t-r_d^t \right)+O\left(( r_d^t-r_u^t )^2\right) \,,
\end{align}
which can become quite significant at small $\tan\beta$.

On the other hand, the RG correction to the iso-scalar coupling  $C_0$  
is, in general, very small. 
From Eq.~\eqref{eq:CppCn}, we see that this coupling combination  gets contributions from $(r_u^t+r_d^t)$ and from $\Delta_s C_s$.
As it was pointed out in Ref.~\cite{DiLuzio:2022tyc},  
in the leading approximation in which only the contribution of the
top-quark Yukawa coupling is kept,  
the combination $(r_u^t+r_d^t)$
is characterized by
a strong cancellation, see for example \eqs{eq:Cu2}{eq:Cd2}, 
and is numerically very small $\sim 0.2\%$.\footnote{From the more accurate numerical 
analysis  in \ref{sec:RGEfit} 
we obtain that for any value of the $m_{\rm BSM}$ scale $|r_u^t+r_d^t|/|r_u^t-r_d^t| \lesssim 0.5\%$. } 
Hence, eventually the leading correction to 
$C_0$ comes from the RG correction $\Delta_s \Delta C_s$ to the last term in \eq{eq:CppCn}. 
It is easy to estimate this contribution from our general results, using $r_s^t=r_d^t$ that holds for universal models.
In the end, we find that the RG corrections to $C_0$ are only about 3\% of the corresponding corrections to $C_3$ and hence this combination of couplings (and the corresponding iso-scalar axion coupling to nucleons, $g_{aN,0}=C_0 m_N/f_a$)
is practically unaffected 
by RG running effects.

\subsection{DFSZ axion couplings beyond tree level}
\label{sec:DFSZ_axion_couplings}

The scalar sector of DFSZ models \cite{Zhitnitsky:1980tq,Dine:1981rt}
features a SM singlet complex scalar   $\Phi$ 
and two Higgs doublets $H_{u,d}$ 
that couple respectively to up- and down-type quarks
in a generation-independent way. 
Under  $\SU(3)_C\times \SU(2)_L\times \U(1)_Y$  
they transform as $\Phi \sim (1,1,0)$, 
$H_u \sim (1,2,-1/2)$ and $H_d \sim (1,2,1/2)$. 
The threefold  re-phasing symmetry of the scalar sector 
$\U(1)_\Phi\times \U(1)_{H_u}\times \U(1)_{H_d}$
is broken to  $\U(1)_{\rm PQ} \times \U(1)_Y$ 
by a renormalizable non-Hermitian operator that can be chosen as 
$H_u H_d \Phi^{\dag 2}$ or $H_u H_d \Phi^{\dag}$.\footnote{The first 
possibility yields a number of domain walls $N_{\rm DW} = 6$ while the 
second $N_{\rm DW}=3$, but they remain otherwise indistinguishable 
from the point of view of low-energy phenomenology.}
There are two possible variants of the model, depending on whether the lepton sector couples to $H_d$ (DFSZ1) 
or to $\tilde H_u = i \sigma^2 H_u^*$ (DFSZ2). For a review see Sec.~2.7.2 in Ref.~\cite{DiLuzio:2020wdo}. 
The Yukawa sector of the DFSZ1 model contains   the following  operators 
\beq
\label{eq:DFSZ1}
\bar q_i u_j H_u \, ,\ \:\bar q_i  d_j H_d\, , \ \:\bar \ell_i e_j H_d\, , 
\eeq
where a sum over generation indices $i,j = 1,2,3$ is left understood, and 
$q_i,\, \ell_i$ denote the quarks and leptons  
$\SU(2)_L$ left-handed (LH) doublets while $u_j,\, d_j,\, e_j$ the right-handed (RH)  singlets. 
The corresponding coefficients for the axion coupling 
at the UV scale $f_a$ are 
\begin{align}
\frac{E}{N} =\frac{8}{3} \, , \ \
C_{u,c,t} (f_a) =\frac{c^2_\beta}{3}  \, ,\ \ C_{d,s,b} (f_a) 
= C_{e,\;\!\mu,\tau} (f_a) =\frac{s^2_\beta}{3}\, , 
\end{align}
with 
$c_\beta \equiv \cos\beta$, 
$s_\beta \equiv \sin\beta$ and 
$\tan\beta = \vev{H_u} / \vev{H_d}\equiv v_u/v_d$. 
The domain in which  
$\tan\beta$ is allowed to vary is obtained by requiring that the 
DFSZ Yukawas remain perturbative up to scales 
of $\mathcal{O}(f_a)$. 
This corresponds to imposing perturbative unitarity on 
Higgs-mediated $2 \to 2$ SM fermion scatterings 
(see e.g.~\cite{DiLuzio:2016sur})
up to $f_a$. 
The perturbative domain is evaluated
by evolving the values of the gauge couplings and of the SM Yukawa couplings 
at $m_Z$ given in Ref.~\cite{Antusch:2013jca} 
up to the scale $m_{\rm BSM}$ employing two-loop RG equations.

\begin{figure}[t!]
\centering
\includegraphics[width=0.45\textwidth]{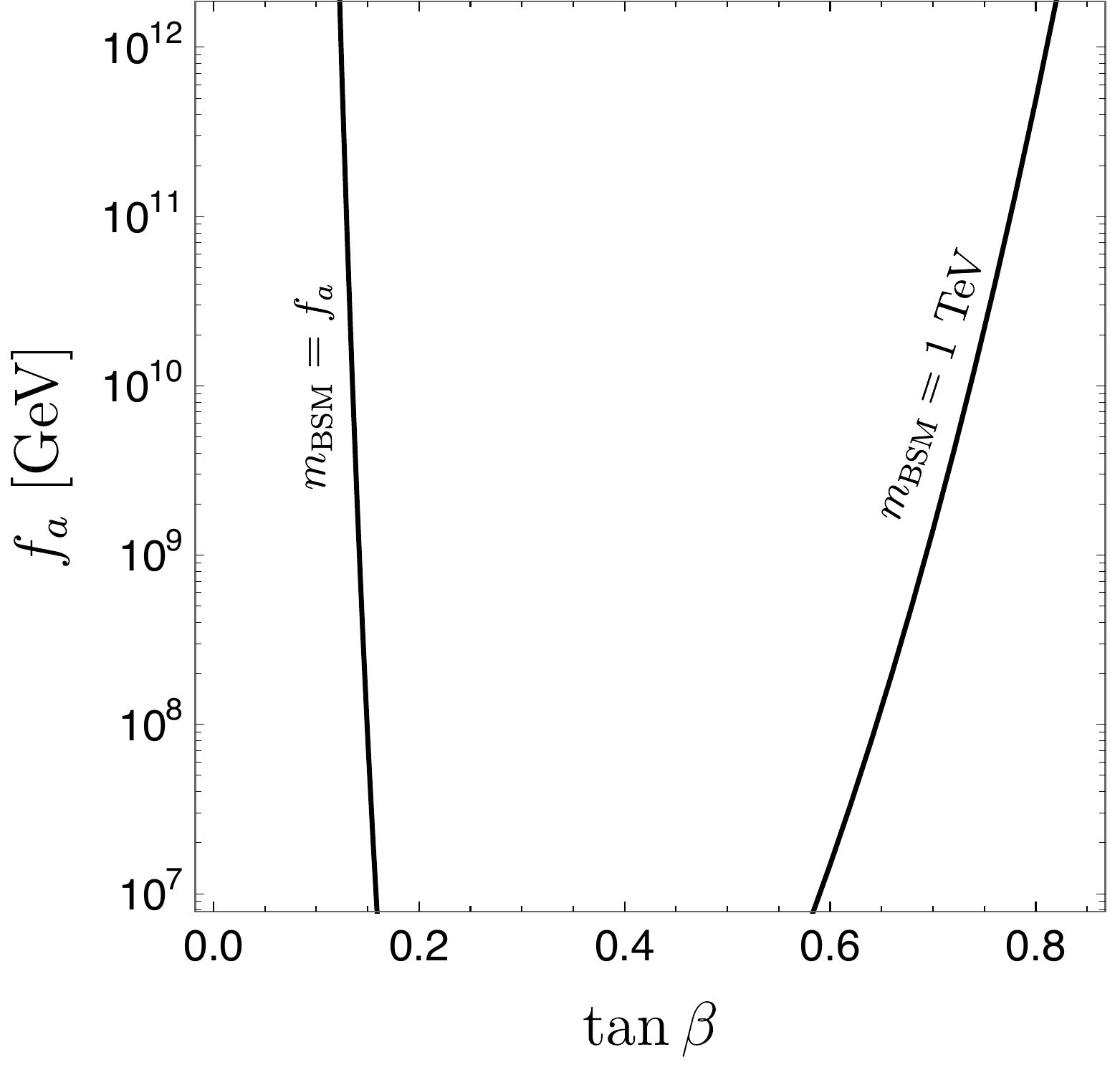}
\caption{$f_a$
dependence of the 
perturbative unitarity 
bounds on $\tan\beta$ at small $\tan\beta$ values.
\label{fig:pertfa}}
\end{figure}

For $m_{\rm BSM } \sim f_a$ the SM Yukawa couplings are RG-evolved from $m_Z$ to $f_a$,  
and upon matching with the DFSZ couplings
$Y^{\rm DFSZ}_t = Y_t(f_a)/s_\beta$ and 
$Y^{\rm DFSZ}_b = Y_b(f_a)/c_\beta$,  by 
requiring $Y^{\rm DFSZ}_{t,b} < \sqrt{16\pi/3}$ \cite{DiLuzio:2020wdo}. 
This yields the perturbative domain
\beq
\label{eq:pertbound1}
\tan\beta \in [0.14, 500] \qquad (m_{\rm BSM} \sim f_a \sim 10^9 \, \text{GeV}) \, ,
\eeq
On the other hand, for $m_{\rm BSM } \ll f_a$ 
one should require a stronger perturbativity constraint on $Y^{\rm DFSZ}_{t,b}$, since running 
effects from $m_{\rm BSM }$ to $f_a$ would tend 
to develop Landau poles in the DFSZ Yukawa couplings 
below $f_a$. 
In this case $Y_{t,b}$ are evolved from $m_Z$ to $m_{\rm BSM}$ within the SM, 
and after matching with the DFSZ couplings
$Y^{\rm DFSZ}_t(m_{\rm BSM}) = Y_t(m_{\rm BSM})/s_\beta$ and 
$Y^{\rm DFSZ}_b(m_{\rm BSM}) = Y_b(m_{\rm BSM})/c_\beta$, 
the running of $ Y^{\rm DFSZ}_{t,b}$ 
from $m_{\rm BSM}$ to $f_a$ is computed in the 
2HDM. 
In the case when $m_{\rm BSM } \sim 1$ TeV 
perturbative unitarity up to $f_a \sim 10^9$\,GeV translates 
in the following interval
\beq
\label{eq:pertbound2}
\tan\beta \in [0.70, 100] \qquad (m_{\rm BSM} \sim 1 \, \text{TeV}) \, .
\eeq
Note that the perturbative domain of $\tan\beta$ 
has a mild (logarithmic) dependence on the PQ scale $f_a$. 
This is shown in \fig{fig:pertfa} 
for the low $\tan\beta$ region 
(a similar dependence is present also for the large $\tan\beta$ region, where running effects are however less important).

The Yukawa sector of the DFSZ2 model contains 
instead
the following  operators 
\beq
\label{eq:DFSZ2}
\bar q_i u_j H_u \, ,\ \:\bar q_i  d_j H_d\, , \ \:\bar \ell_i e_j \tilde H_u\, ,   
\eeq
and the corresponding axion coupling coefficients are 
\begin{align}
\!\frac{E}{N} = \frac{2}{3} \, , \ \,
C_{u,c,t} (f_a) =
-C_{e,\;\!\mu,\tau} (f_a) =
\frac{c^2_\beta}{3}\, ,\ \, C_{d,\,s,\,b} (f_a) =\frac{s^2_\beta}{3}\, , 
\end{align}
with $\tan\beta$ defined in the same perturbative domain as in DFSZ1. 

\begin{table*}[t!]
\begin{center}
\begin{tabular}{ |l|l|l| } 
\hline
 Coupling (DFSZ1) & Coupling (DFSZ2) & Approximate Correction   \\[0.3ex] \hline  
$ C_0 \simeq -0.20 $  & $ C_0 \simeq -0.20 $  &$\Delta C_0 \approx 0$ \\ [0.3ex]  
 $C_3  \simeq -0.43\,\sin^2\beta $  & $C_3  \simeq -0.43\,\sin^2\beta $ & $\Delta C_3 \simeq -0.12 \, l(x) \,\cos^2\beta $   \\[0.3ex] 
$C_e =\frac13 \sin^2\beta $ & $C_e =-\frac13 \cos^2\beta $ & $\Delta C_e\simeq 0.094  \, l(x) \,\cos^2\beta $    \\ [0.5ex] 
 $C_\gamma =\frac83 -1.92 $ &  $C_\gamma =\frac23 -1.92 $ & $\Delta C_\gamma= 0  $   \\ [0.5ex] 
 \hline 
\end{tabular}
\end{center}
    \caption{
    RG corrections (third column) to the DFSZ1 and DFSZ2 couplings 
    (listed respectively in the first and second column) 
    in the approximation of 
    keeping only the contribution from the top Yukawa coupling $Y_t$. 
    The corrections are given in terms of $\beta$ and of 
    $l(x) = \ln(\sqrt{x}-0.52)$, where 
    $x=\log_{10}(m_{\rm BSM}/{\rm GeV})$ 
    parameterizes the new physics scale.  
    While the corrections to the quark couplings in DFSZ1/2 are the same, 
    for the leptons the relative corrections differ: 
    $\Delta C_e/C_e \simeq \cot^2\beta$ (DFSZ1) and 
    $\Delta C_e/C_e \simeq {\rm const.}$ (DFSZ2).
    \label{tab:DFSZ_corrections}
    }
\end{table*}

\begin{figure*}[t!]
\centering
\includegraphics[width=0.45\textwidth]{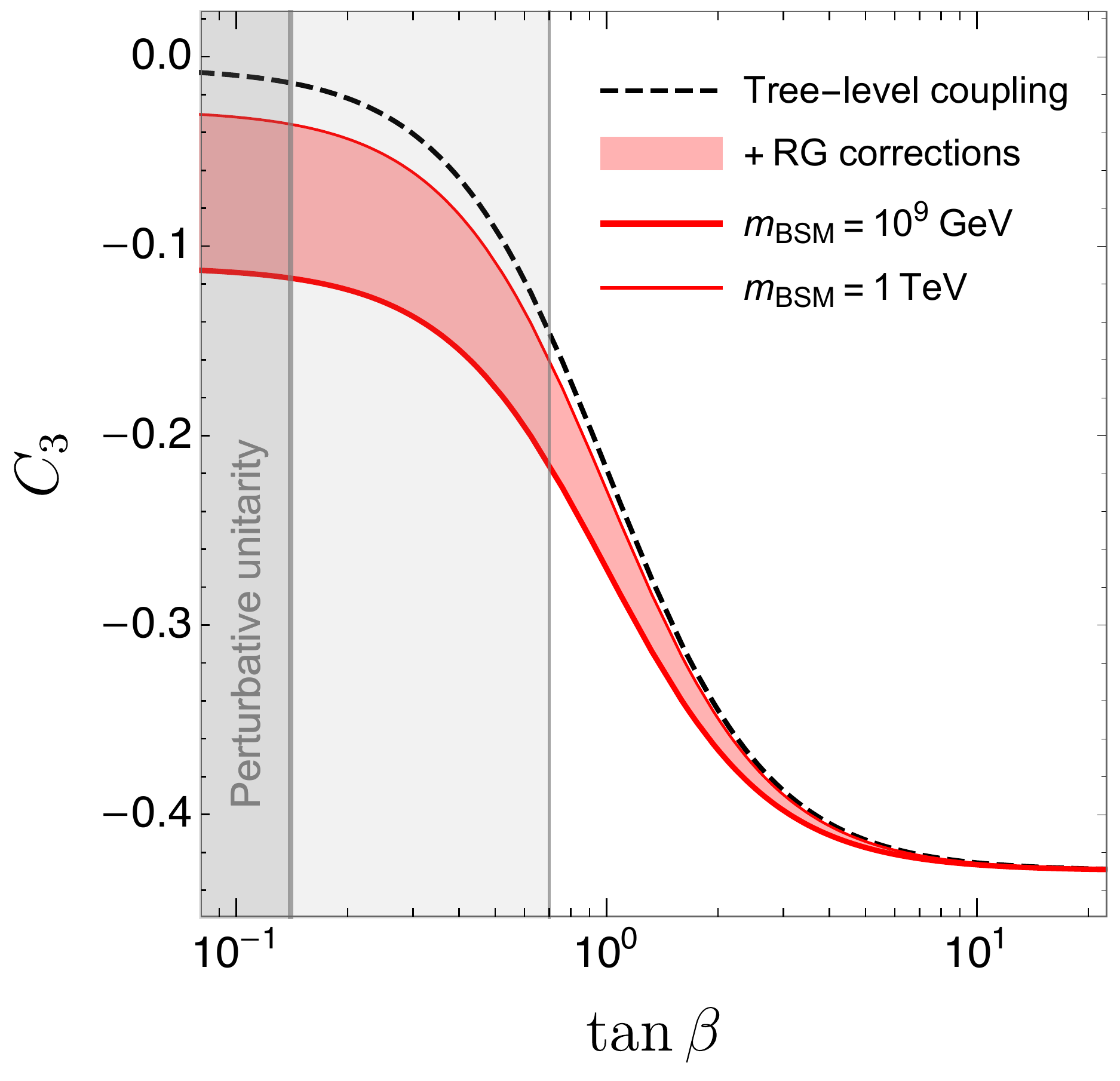}
\qquad
\includegraphics[width=0.45\textwidth]{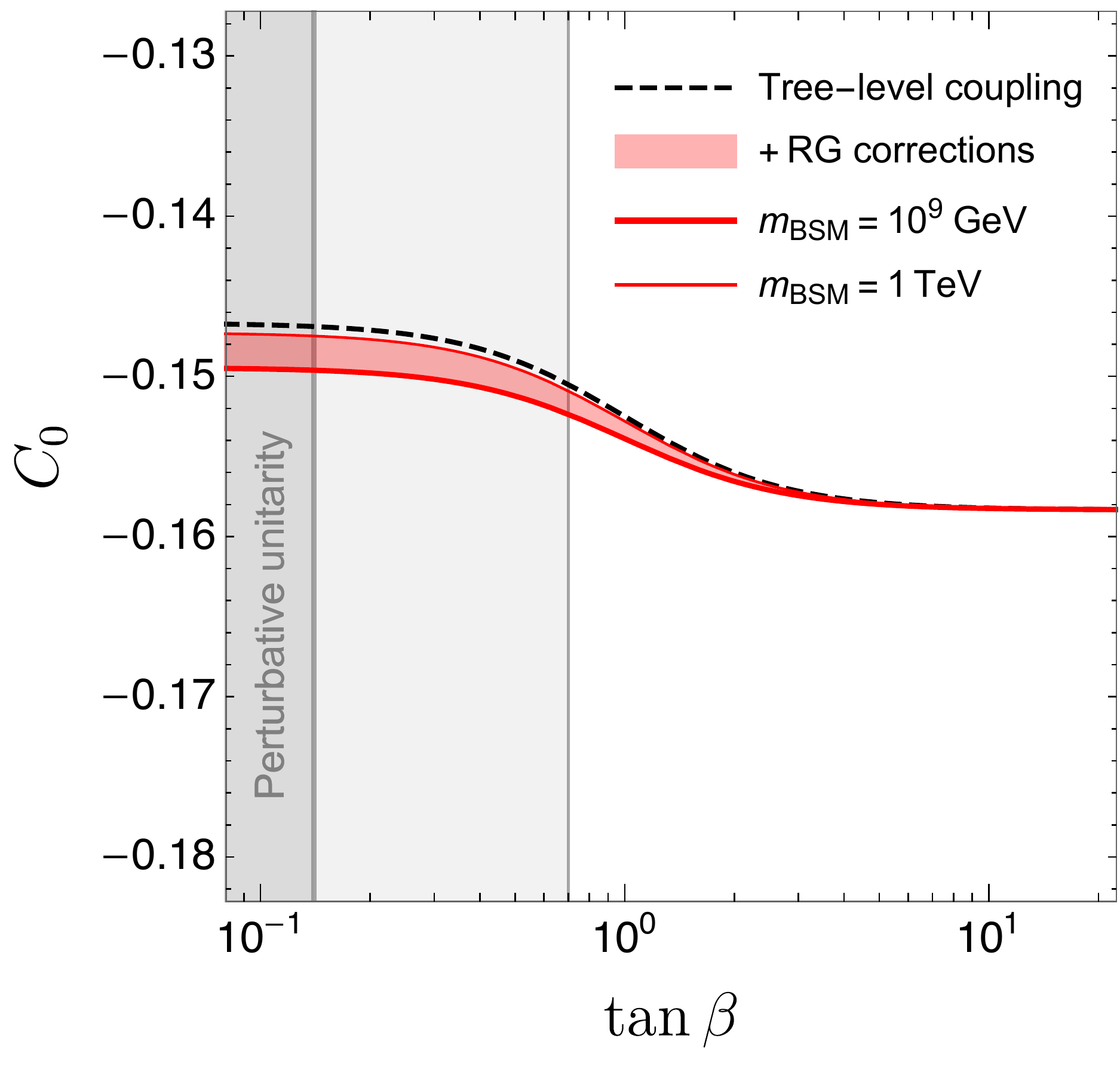}
\caption{Running axion coupling 
combinations ($C_3$ and $C_0$) in DFSZ
as a function of $\tan\beta$. 
The red band encompasses  the range of the corrections 
for $m_{\rm BSM}/\text{GeV} \in [10^3, 10^9]$. 
Perturbativity bounds on $\tan\beta$ also depend on 
$m_{\rm BSM}$, 
with the thick (thin) grey line corresponding to 
$m_{\rm BSM} = 10^{9}$ GeV 
(1 TeV).
\label{fig:Carunning}}
\end{figure*}

Let us now proceed to discuss the impact of 
running effects in the DFSZ model. 
Approximate RG corrections to 
the axion couplings are collected in Tab.~\ref{tab:DFSZ_corrections}. 
Note that in the case of DFSZ1 the iso-vector combination $C_3$, 
as well as the 
axion-electron coupling
$C_e$, receive large  corrections 
at small $\tan\beta$, that is when the tree-level coupling vanishes.\footnote{Note that the suppression 
of $C_3$ at small 
$\tan\beta$ arises due to an accidental cancellation 
between $C_u-C_d$ and the quark mass term, cf.~\eq{eq:CpmCn}.} 
On the other hand, the RG corrections on the 
iso-scalar combination $C_0$ remain small in the whole 
$\tan\beta$ range. 
This is also displayed in \fig{fig:Carunning}, 
where the dashed line corresponds to the 
tree-level result, while the red band encodes the range 
of RG corrections obtained by varying $m_{\rm BSM}$ between
$f_a = 10^9\,$GeV (lower border of the red region) and 
$1\,$TeV (upper border of the red region). 
We see that although for $m_{\rm BSM} = 1\,$TeV
the running couplings trace closely the tree-level couplings as long as $\tan\beta > 1$, also in this case 
RG corrections become non-negligible at small $\tan\beta$.

\section{RG effects on QCD axion phenomenology}
\label{sec:runningpheno}

\begin{figure*}[t!]
\centering
\includegraphics[width=0.32\textwidth]{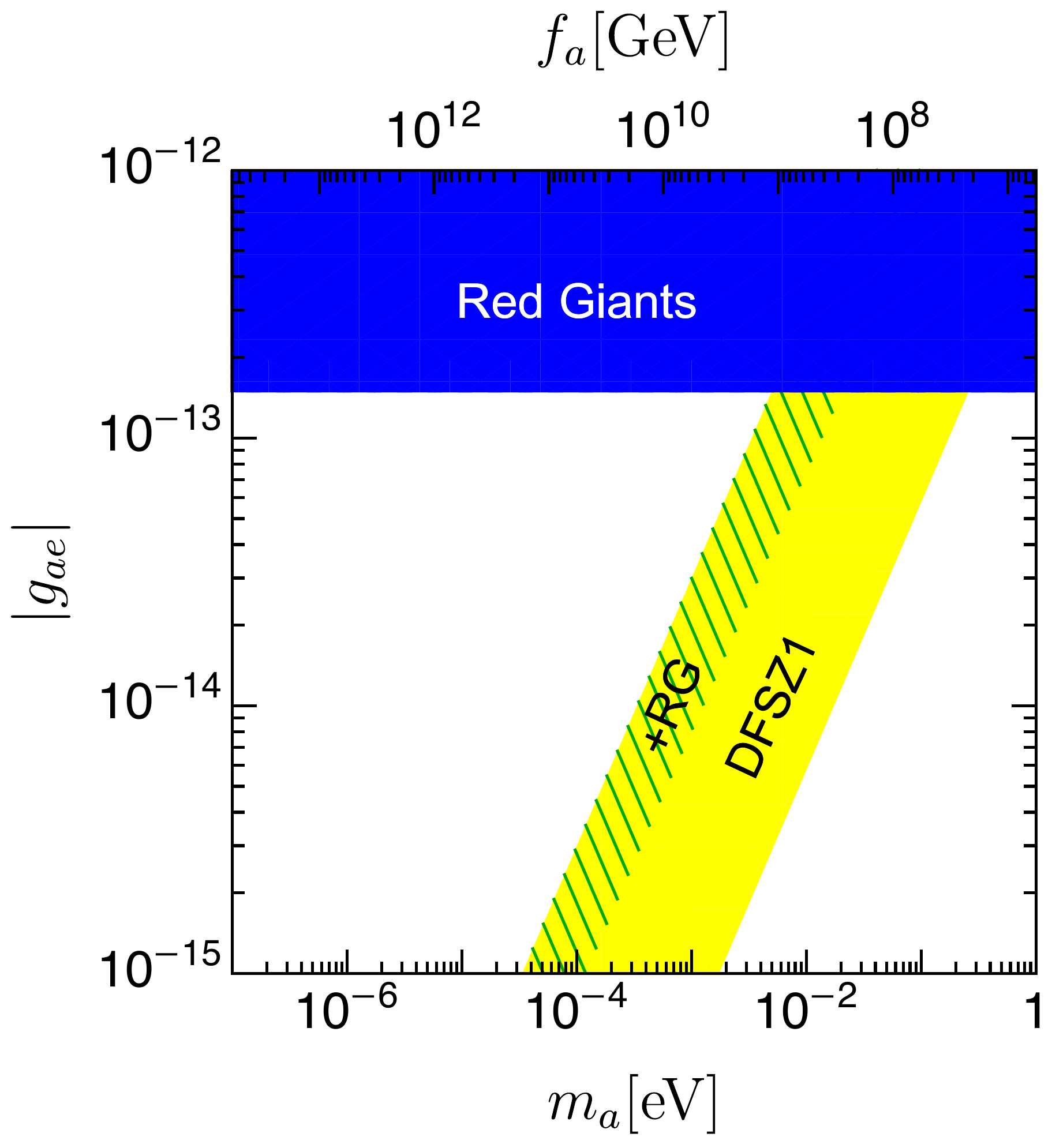}
\ \ 
\includegraphics[width=0.32\textwidth]{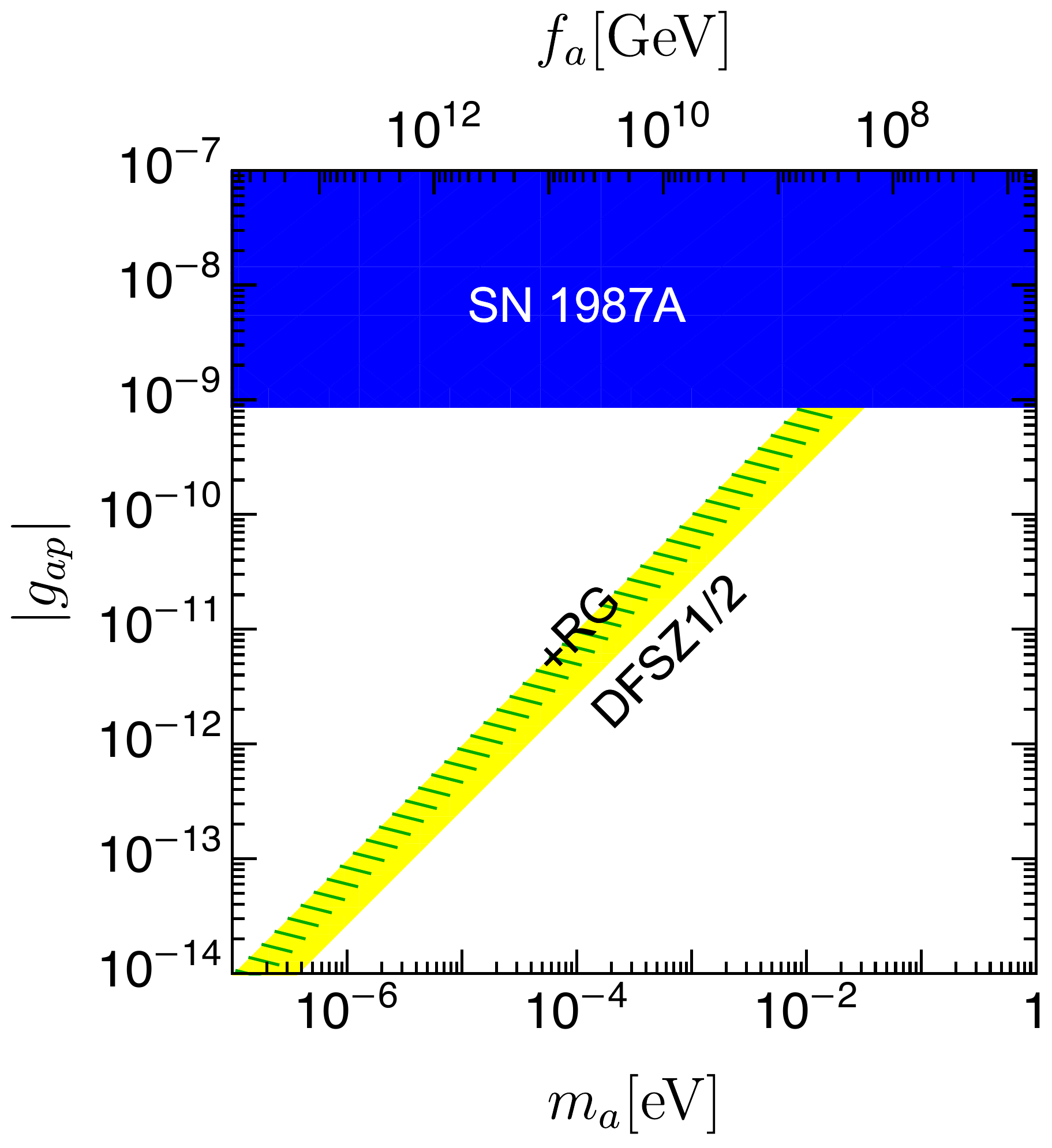}
\ \ 
\includegraphics[width=0.32\textwidth]{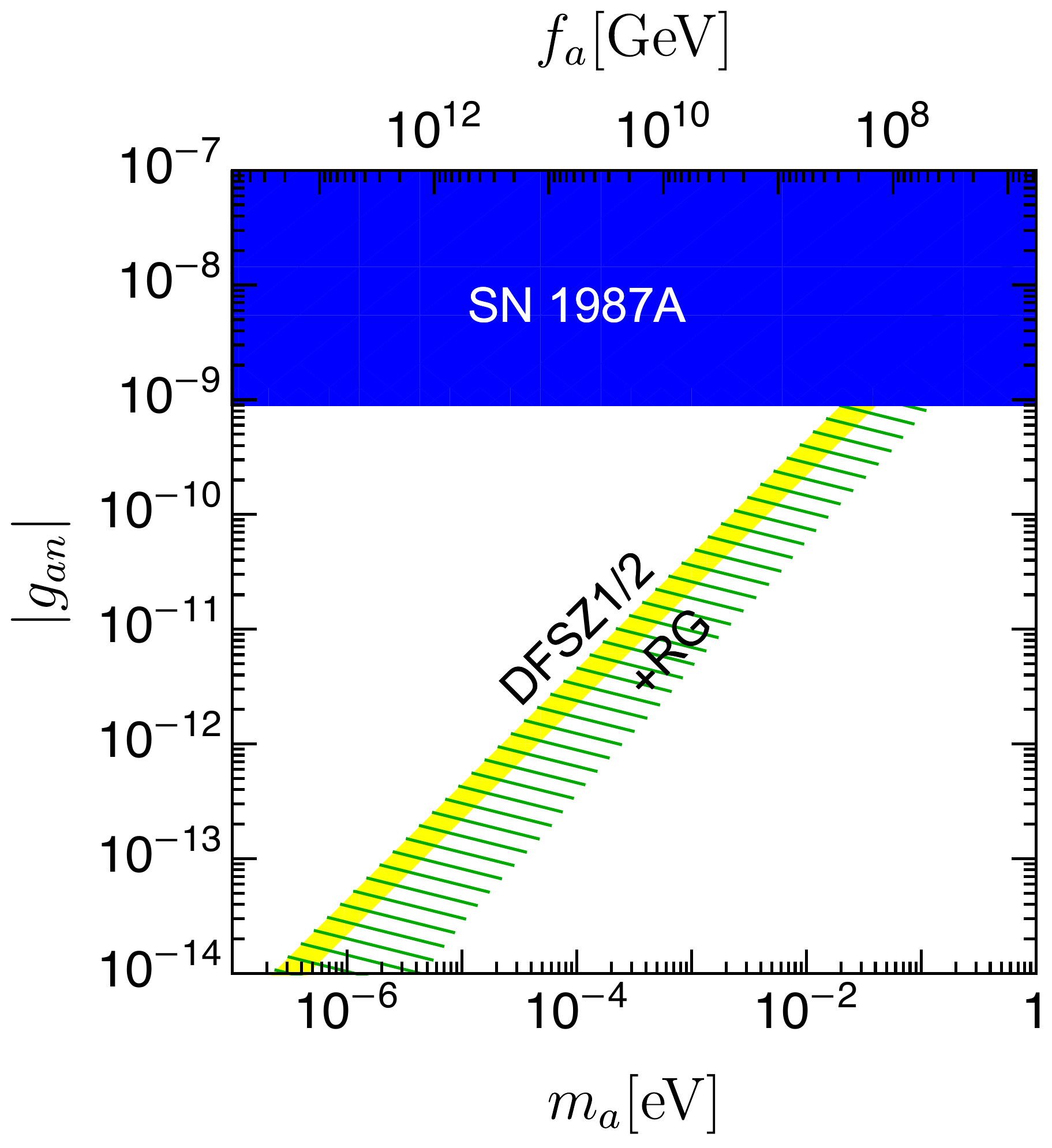}
\caption{
Redefinition of the predicted bands for 
the DFSZ couplings $g_{ae}^{\rm DFSZ1}$  (left), $g_{ap}^{\rm DFSZ1/2}$  (middle), and $g_{an}^{\rm DFSZ1/2}$ (right) induced by RG effects (hatched green region). 
The size of the band correspond the perturbativity  unitarity bounds on $\tan\beta$ (see text). The $m_{\rm BSM}$ new physics scale is set to the maximal value
$m_{\rm BSM}=f_a$.
\label{fig:DFSZ_bands}}
\end{figure*}

In the following section, 
we will discuss the phenomenological impacts of the RG corrections.
In particular, we will see how these affect axion astrophysical and cosmological bounds, as well as the  sensitivity of terrestrial experimental searches. 
For clarity, we will refer to the DFSZ axion models, 
even though our results can be applied to other axion models.

Several observables depend dominantly 
(or entirely) on $C_3$ and thus,
as discussed in Sect.~\ref{sec:runningcoupl}, are subjected to large RG-induced modifications. 
These include for example the coupling to pions, which is responsible for the axion thermalization in the early Universe (via $ \pi \pi \leftrightarrow \pi a $), which controls the hot dark matter (HDM) bound 
discussed in Sect.~\ref{sec:thermal_axions}.
More recently, it has also been shown that the 
pion-nucleon scattering
may be responsible for a large contribution to the axion emission rates in dense media, particularly in SNe~\cite{Carenza:2020cis,Fischer:2021jfm,Lella:2022uwi}. 
Finally, the iso-vector coupling $C_3$ is entirely responsible for the nuclear reaction process 
$p+d\to\, ^3{\rm He}+a$,
which is one of the most efficient and widely studied production mechanisms of axion-like 
particles from solar nuclear reactions~\cite{Raffelt:1982dr,CAST:2009jdc,Borexino:2012guz,Bhusal:2020bvx,Lucente:2022esm}.
On the other hand, the nucleon coupling to $^{57}$Fe, 
relevant for axion production through 
nuclear de-excitations in the Sun \cite{Raffelt:1982dr,CAST:2009jdc,DiLuzio:2021qct}, 
turns out to be less 
sensitive to RG corrections (see Tab.~\ref{tab:RG_corrections}).

The axion-electron coupling $C_e$, which plays a significant role in astrophysics (see Section~\ref{sec:astrophysics}) as well as  in terrestrial experimental searches (see Section~\ref{sec:experiments}), is also subjected to large RG corrections. 
In fact, $r_e^t\approx r_d^t$.\footnote{This conclusion is based on the same arguments presented above, where it was argued that $r_u^t+r_d^t\simeq 0$. Deviations from these relations are due to subleading contributions of Yukawa terms other than $Y_{t}$. A detailed discussion can be found in 
Section~3 of Ref.~\cite{DiLuzio:2022tyc}.  
}
Thus, using $r_u^t+r_d^t\approx 0$ and Eq.~\eqref{eq:r_3}, we find
\begin{align}
\label{eq:r_e}
r_e^t\simeq -\frac12 \left( r_u^t-r_d^t \right) \approx 0.27 \ln \left( \sqrt{x} - 0.52 \right)\,,
\end{align}
where $x=\log_{10} \left( m_{\rm  BSM}/{\rm GeV} \right)$ parameterizes the new physics scale.\footnote{This result should not be surprising since it holds in the limit of $|r_u^t+r_d^t|\approx 0$ and, as discussed above, $|r_u^t+r_d^t|/|r_u^t-r_d^t|\lesssim 0.5\%$.}
Therefore the corrections to  $C_e$  can also be expressed in terms of $\Delta C_3$.
Specifically, from our previous results we see that 
$\Delta C_e=-0.5 \,C_t(f_a) \, (r_u^t-r_d^t)=-0.78\, \Delta C_3$.
Thus, at our level of approximation all running effects can be expressed in terms of $\Delta C_3$, which for DFSZ axions is $\propto \cos^2\beta$.
A general consequence of this observation is that, in the case of DFSZ axions, the running effects are mostly relevant only at small $\tan \beta$, something that is apparent from our numerical analysis.\footnote{The only exception is in cases where the $\cos\beta$ dependence cancels, as for $\Delta C_e/C_e$ in the DFSZ2 model (see the Red Giant bound on DFSZ2 axion in the right panel of Fig.~\ref{fig:DFSZ_astro}).}

Before moving to the phenomenological study, 
it is instructive to anticipate how RG corrections 
will modify the usual DFSZ bands, obtained by varying the value of $\tan\beta$ within the perturbative unitarity limits,  for the   $g_{ae},\, g_{ap},\, g_{an}$ couplings 
(see for example the section on axions in the Review of Particle Physics~\cite{ParticleDataGroup:2022pth}). 
This can be easily estimated by considering the  RG corrections to the electron and to the nucleon couplings $C_{p,n} = C_0 \pm C_3$ 
given in Table~\ref{tab:DFSZ_corrections}. 
The results are shown in Fig.~\ref{fig:DFSZ_bands},
where we have taken $m_{\rm BSM} = f_a \propto 1/m_a $ to maximize the RG effects. In each panel  the bands correspond to varying $\tan\beta$ in the interval $[0.14,100]$.
The first figure shows the modification of the usual
$g_{ae}$ band in DFSZ1.  For this case we obtain the most dramatic effect, that is a marked reduction of the width of the band that, after including RG corrections, shrinks down to the green hatched region.
This is due to the fact that the  tree-level suppression of $g_{ae}^{\rm DFSZ1}$ in the limit $\sin\beta \to 0$ is cut-off at small $\tan\beta$ by the RG correction proportional to $\cos^2\beta$.  Instead there is 
not such a dramatic effect for $g_{ae}^{\rm DFSZ2}$   since both the tree level coupling and the RG correction are proportional to $\cos^2\beta$. 
In the second panel in Fig.~\ref{fig:DFSZ_bands}
we show the RG effect on the band for $g_{ap}$ 
(that is the same in DFSZ1 and in DFSZ2). In this case we see that the shrinking of the allowed band is much less pronounced. 
Finally, the third panel shows the RG effects on the 
$g_{an}$ band. In this case the allowed band is sizeably widened, which this is due to a cancellation 
in the $\tan\beta$ independent part of the coupling, 
which enhances the overall dependence on this parameter.

\begin{table*}[t!]
\begin{center}
\begin{tabular}{ |l|l|l| } 
\hline
 Coupling & Approximate Correction & Processes  \\[0.3ex] \hline  
$ C_0 $  & $\Delta C_0 \approx 0$ &   \\ [0.7ex] 
 \multirow{2}{10em}{$C_3$} & \multirow{2}{11em}{$\Delta C_3 \simeq 0.64 \,C_t(f_a)\, (r_u^t-r_d^t)$} &  Axion thermalization: $ \pi \pi \leftrightarrow \pi a $~\cite{Chang:1993gm,Hannestad:2005df,DiLuzio:2021vjd,Notari:2022zxo,DiLuzio:2022gsc} \  
 \\ 
 &  & Deuteron processes: $p+n\leftrightarrow d+a$~\cite{Borexino:2012guz,Bhusal:2020bvx,Lucente:2022esm}\\ [0.7ex] 
$ C_p=C_0+C_3 $  & $\Delta C_p \approx \Delta C_3$ & Astrophysics/experiments~\cite{DiLuzio:2021ysg,Irastorza:2018dyq,Sikivie:2020zpn}  \\ [0.7ex] 
$ C_n=C_0-C_3 $  & $\Delta C_n \approx -\Delta C_3$ & Astrophysics/NMR experiments~\cite{Arvanitaki:2014dfa,Garcon:2017ixh}  \\ [1.8ex] 
$C_{\rm SN}\simeq 1.4\left( C_0^2+0.11\,C_0C_3 +1.3\, C_3^2 \right)^{1/2}\,,
$ & $\Delta C_{\rm SN}\simeq \left( 2.5\,C_3+0.11\,C_0 \right)\frac{\Delta C_3}{C_{\rm SN}}\,.$ 
& SN 1987A bound~\cite{Lella:2022uwi} \\  [2.2ex] 
 $C_{\rm Fe}  =  C_0-0.77\,C_3 $ & $\Delta C_{\rm Fe}=
 -0.77 \Delta C_3$  &  Axion production/detection in $^{57}$Fe~\cite{Raffelt:1982dr,CAST:2009jdc,DiLuzio:2021qct}\\  [0.9ex] 
\multirow{2}{10em}{$C_e  $} & \multirow{2}{10em}{$\Delta C_e=-0.78\, \Delta C_3$}  &  Axion production in stars~\cite{DiLuzio:2021ysg} \\ [0.3ex] 
 &   &  Axion detection (Xenon etc.)~\cite{LUX:2017glr, PandaX-II:2020udv, XENON:2022ltv} \\  [0.9ex] 
\multirow{2}{10em}{$C_\gamma$} & \multirow{2}{10em}{$\Delta C_\gamma= 0$}  &  Axion production in stars and labs~\cite{DiLuzio:2020wdo}  \\ [0.3ex] 
 &   &  Most axion detection experiments~\cite{Irastorza:2018dyq,Sikivie:2020zpn}  \\  [0.9ex] 
 $C_{\rm hel}=\left[ C_{\gamma}^2\left( C_{\gamma}^2 +(37 C_e)^2 \right) \right]^{1/4}$ & $\Delta C_{\rm hel}
 =\dfrac{C_{\rm hel}}{2} 
 \left( \dfrac{\Delta C_e/C_e}{1+\left( {C_\gamma}/{(37 C_e)}  \right)^2} \right)$
  &  Axion coupling for Sikivie helioscopes~\cite{Sikivie:1983ip,DiLuzio:2020wdo} \\  [3.8ex] 
 \hline
\end{tabular}
\end{center}
    \caption{
    RG corrections (second column) to axion couplings listed in the first column, 
    in the approximation of keeping only the contribution from the top Yukawa coupling $Y_t$. 
    Note that in this approximation all the various corrections can be expressed just in terms of $\Delta C_3$ 
    given in the second line, with $r^t_u - r^t_d$ given in Eq.~\eqref{eq:r_3}.
    \label{tab:RG_corrections}
    }
\end{table*}

The most important phenomenological consequences of 
the RG corrections to the axion couplings
will be analyzed in the following sections.

\subsection{Astrophysical constraints} 
\label{sec:astrophysics}

In this Section, we discuss the impact of RG corrections to astrophysical observables. 
For reference, we will 
mostly focus on the DFSZ1 axion model. 
The analysis for DFSZ2 goes along similar lines.

Axions can be copiously produced in stars, mostly due to thermal processes (see Ref.~\cite{DiLuzio:2021ysg} for a recent review).
Here we will not consider astrophysical bounds on the axion-photon coupling~\cite{Ayala:2014pea,Straniero:2015nvc,Giannotti:2015kwo,Giannotti:2017hny}, since $C_{\gamma}$ does not receive any relevant RG correction.
We focus instead on the axion-electron and on the axion-nucleon couplings. 

\begin{figure*}[t!]
\centering

\includegraphics[width=0.45\textwidth]{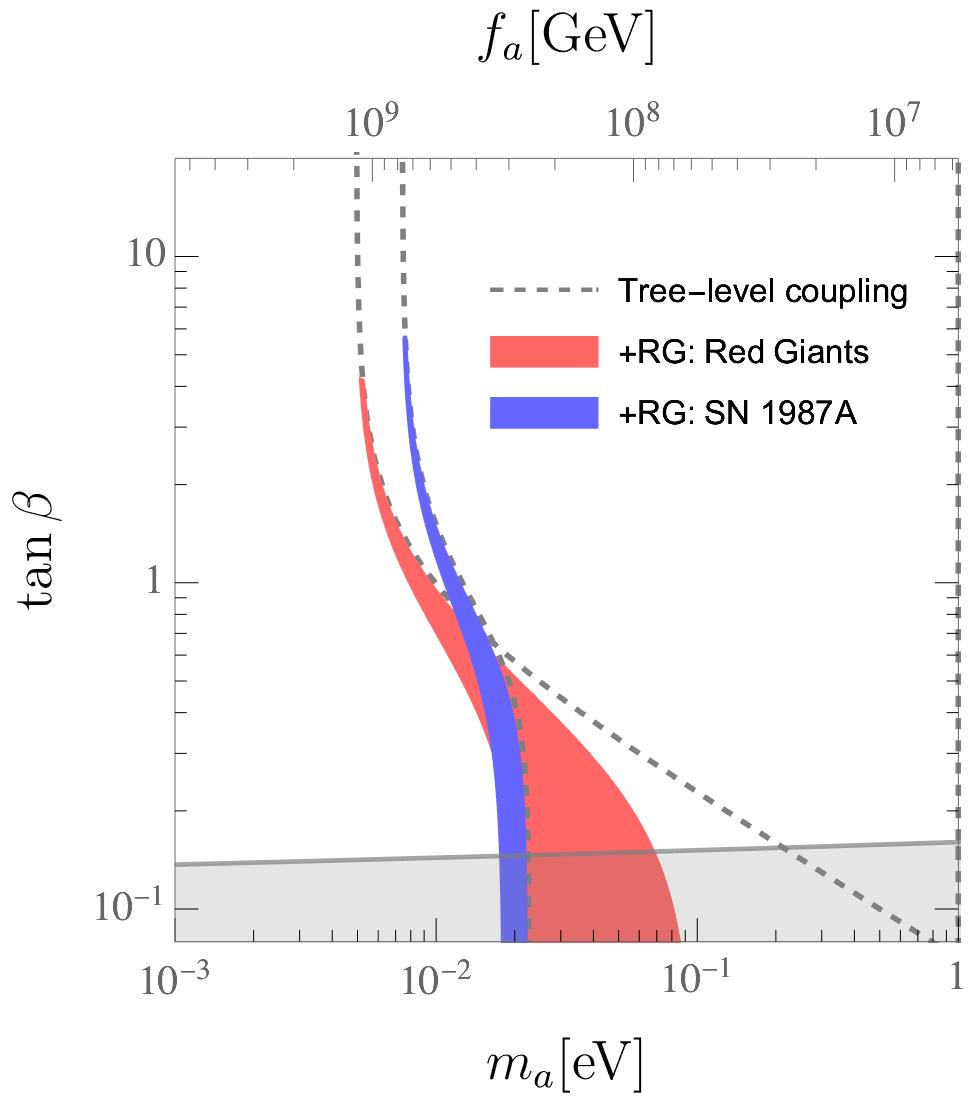}
\qquad 
\includegraphics[width=0.45\textwidth]{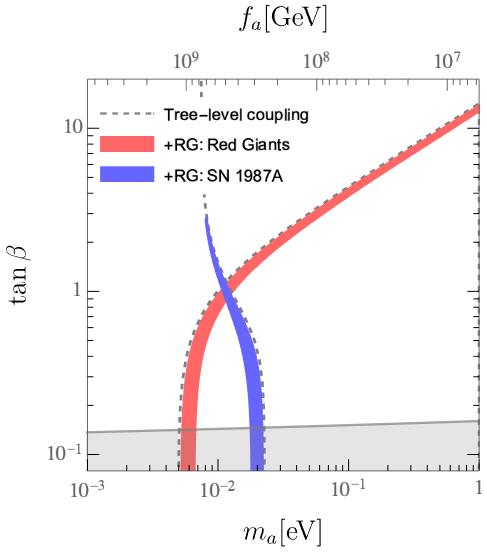}
\caption{RG effects on astrophysical  axion bounds from Red Giants (red bands) and SN1987A (blue bands) for the DFSZ1 (left panel) and DFSZ2 (right panel) models,  compared to the tree level results (black dashed lines).
The gray line corresponds to the perturbative 
unitarity bound on $\tan\beta$ 
for $m_{\rm BSM} = f_a$.
\label{fig:DFSZ_astro}}
\end{figure*}

The most stringent astrophysical bound on the axion-electron coupling is derived from observations of the tip of the red giant branch (RGB) in globular clusters. 
The production of axions during the RGB evolution cools the core, playing a role similar to that of neutrinos, and thus delays the helium ignition.
The delay leads to a larger helium core and, consequently, to a higher stellar luminosity. 
Thus, comparison between observations and predictions for the luminosity of the RGB tip (the brightest stars in the RGB) is an efficient way to test anomalous channels of stellar cooling.
The most recent analysis has set the constraint $|g_{ae}|\leq 1.48\times 10^{-13}$ (at 95\% C.L.) ~\cite{Straniero:2020iyi}.
From the definition of this coupling, $g_{ae}=C_{e} m_e/f_a$,
the RGB bound translates into
\begin{align}
\label{eq:RGB_bound}
|C_{e}|\leq 1.65\times 10^{-3}\left( \frac{ m_a}{\rm eV} \right)^{-1} \,.
\end{align}
This relation provides an upper bound for the axion mass at any given value of 
$\tan\beta$ and $x=\log_{10} \left( m_{\rm  BSM}/{\rm GeV} \right)$.
The full numerical results are shown in 
Fig.~\ref{fig:DFSZ_astro}, 
where the red-shaded bands incorporate the range fixed by the possible values of $m_{{\rm BSM}} \in [1 \ \text{TeV}, f_a]$.

We can gain some intuition about these effects using our approximate results, shown in Table.~\ref{tab:DFSZ_corrections}.
In the case of the DFSZ1 model (left panel 
in Fig.~\ref{fig:DFSZ_astro})
we can conveniently rewrite the RGB bound on the axion mass as follows\footnote{Notice that in the range of $m_{\rm BSM}$ we are considering here, the absolute value in Eq.~\eqref{eq:RGB_bound_Ce} is unnecessary.}
\begin{align}
\label{eq:RGB_bound_Ce}
\left( \frac{ m_a}{\rm eV} \right)  \leq
\frac{1.65\times 10^{-3}}
{\Big |
\left( \frac{1}{3} -0.094  \,l(x) \right)\sin^2\beta +0.094\,l(x) \Big |   }\,,
\end{align}
where $l(x)=\ln \left( \sqrt{x} - 0.52 \right)$.
The first important observation is that in the limit $l(x)\to 0$, that is ignoring the RG corrections, the RGB bound on the mass is a monotonic function of $\tan\beta$ and disappears in the limit of small $\tan \beta$. 
This result is modified by the RG corrections, which 
in the limit $\tan\beta\to 0$ still 
provide a useful limit on the axion mass,  $m_a\leq 0.018\,{\rm eV}/l(x) $. 
From these considerations, we can conclude  that the RG correction to the RGB bound becomes particularly important in the low $\tan \beta$ limit, a result confirmed by the full numerical result shown in Fig.~\ref{fig:DFSZ_astro}.
The most conservative value for the RGB bound 
corresponds to $m_{\rm BSM}=1\,$TeV, for which we obtain, in our approximation,  
$m_a\leq 0.018\,{\rm eV}/l(3)\simeq 8.75\times 10^{-2}\,$eV,
which agrees well with the complete numerical result shown in the left panel of Fig.~\ref{fig:DFSZ_astro}.
In the case of the DFSZ2 model instead, there is not such a striking feature, and this is because in this case both the coupling $g_{ae}$ and its RG correction depend on $\cos\beta$. The RGB bound for this  case is shown in the right panel of Fig.~\ref{fig:DFSZ_astro}.

Let us now move to the axion-nucleon coupling and analyze 
the axion bound from SN1987A.\footnote{SN1987A is not the only astrophysical probe of the axion-nucleon 
coupling. 
Neutron stars also provide  strong bounds (see, e.g., Ref.~\cite{Buschmann:2021juv}). 
However, the bound from SN1987A is the most discussed in the literature, and thus it provides a good example for the impact of RG corrections.}
This is a quite more complex problem since  axion production in a SN environment, at temperatures of order $\sim 30\,{\rm MeV}$ and 
densities in excess of $\sim10^{14}\,{\rm g/cm^3}$, gets contributions also from pions~\cite{Carenza:2020cis,Fischer:2021jfm,Choi:2021ign} and  from  $\Delta$ baryon resonances~\cite{Ho:2022oaw}. 
Following the notation of Ref.~\cite{Lella:2022uwi}, 
the effective low-energy axion-nucleon interaction relevant for the axion production processes in a SN environment is 
given in \eq{eq:interactions}. 
The second term in the first line 
describes the usual axion-nucleons interactions. 
The third line contains the axion-pion interactions.
The four-particle interaction vertex in the third line 
accounts for the pion-nucleon contact term 
recently discussed in Ref.~\cite{Choi:2021ign}, 
while the last line accounts for 
the axion couplings to the $\Delta$-resonances, whose
contribution to the axion emissivity has been recently calculated 
in Ref.~\cite{Ho:2022oaw}. 
The interaction Lagrangian in Eq.~\eqref{eq:interactions} can be used to compute the axion emissivity due to  nucleon-nucleon ($NN$) bremsstrahlung, $NN\to NNa$, as well as  the Compton-like pion scattering processes, $\pi^{-} p\to n a$,  including also the contribution from the $\Delta$ resonances (see Ref.~\cite{Lella:2022uwi} for an updated overview).

In general, 
the axion luminosity from a SN,  $L_a$, depends  only on a particular combination of $C_0$ and $C_3$, and thus only this combination can be constrained. 
The luminosity can be expressed as \cite{Lella:2022uwi}
\begin{align}
\label{eq:SN_Luminosity}
L_a=\epsilon_0 \left( \frac{m_N}{f_a} \right)^2 C_{\rm SN}^2\times 10^{70}\,{\rm erg/s}\,,
\end{align}
where $\epsilon_0$ is a numerical factor and 
\begin{align}
\label{eq:CSN_general}
C_{\rm SN}=a\left( C_0^2+b \,C_3^2+c\, C_0 C_3 \right)^{1/2}\,.
\end{align}
The numerical values of the coefficients $\epsilon_0$, $a$ and $b$ 
can be found in Table~\ref{tab:SN_luminosity}.
In the table we present, in the first line, the results obtained by considering only the purely $NN$ bremsstrahlung production, which corresponds to the first line of Eq.~\eqref{eq:interactions}.  
The results  for the total emission rate are given in the second line (we remind to the reader that up until very recently, the $NN$ bremsstrahlung production was the only process considered for  estimating SN axion emission rate.) 

\begin{table}[]
\centering
\begin{tabular}{|l|c|c|c|c|c|}
\hline
      & $\epsilon_0$ & $a$ & $b$ & $c$   & $\bar m$ [meV] \\ \hline
$NN$  & 2.42         & 1.5 & 0.5 & -0.36 & 6.4            \\ 
Total & 3.86         & 1.4 & 1.3 & 0.11  & 5.0            \\ \hline
\end{tabular}
\caption{Parameters for the axion luminosity from SN entering Eqs.~\eqref{eq:SN_Luminosity} and \eqref{eq:CSN_general}.
The coefficients are calculated at a post-bounce time of 1s (see Ref.~\cite{Lella:2022uwi}).
The first row refers to the $NN$ bremsstrahlung contribution only~\cite{Carenza:2019pxu}, 
ignoring the pion scattering processes and the $\Delta$ resonance contribution.
The second row gives the total contribution, calculated from the results in Ref.~\cite{Lella:2022uwi}.
The mass parameter $\overline m$ is 
defined in \eq{eq:SN87A_bound}.
}
\label{tab:SN_luminosity}
\end{table}

Notice that, as evident from Table~\ref{tab:SN_luminosity}, the addition of the pion-induced scatterings increases 
the relative importance of $C_3$ (controlled by the coefficients $b$ and $c$) 
and thus enhances the effects of the RG corrections. 
More specifically, from Eq.~\eqref{eq:CSN_general}, and assuming $\Delta C_0\approx 0$, we find 
\begin{align}
\label{eq:}
\Delta C_{\rm SN}\simeq a^2\left( b\,C_3+\frac{c}{2}\,C_0 \right)\frac{\Delta C_3}{C_{\rm SN}}\,.
\end{align}
As expected, the RG effects are reduced in the case of purely NN-bremsstrahlung production, due to the partial cancellation between the $b$ and $c$ terms.\footnote{We should be cautious, however, since this expression is valid only in the limit of $\Delta C_3/C_{\rm SN}\ll 1$ and this condition is not always met.  In particular, it is violated at low $\tan\beta$ and high $m_{\rm BSM}$.}

Imposing $L_a\leq L_\nu=3\times 10^{52}\,{\rm erg/s}$~\cite{Lella:2022uwi}, we find the bound on the axion mass
\begin{align}
\label{eq:SN87A_bound}
m_a\leq \frac{\bar m}{C_{\rm SN}}\,,~~ {\rm with~~~~} 
\bar m =\frac{9.9\,{\rm meV}}{\sqrt{\epsilon_0}} \, .
\end{align}
In the case of for DFSZ axions, this bound is shown in Fig.~\ref{fig:DFSZ_astro}.

Specializing  on the DFSZ axion case,  we immediately find from 
Tab.~\ref{tab:DFSZ_corrections}, 
\begin{align}
\label{eq:CSN_DFSZ}
C_{\rm SN}^{\rm DFSZ}=0.2 \,a\sqrt{1+2.15 \,c\sin^2\beta + 4.5\,b\sin^4\beta} \, .
\end{align}
and 
\begin{align}
\label{eq:SN_correction}
\left(\frac{\Delta C_{\rm SN}}{C_{\rm SN}}\right)^{\rm DFSZ} = 
\left[ \frac{\left( 0.30\,c +1.3\,b\sin^2\beta \right)\cos^2\beta}
{1+2.15\,c\sin^2\beta + 4.6\,b\sin^4\beta}  \right]
\,l(x) \, .
\end{align}
Note that the above expression is never larger than about 10\%-15\%. 
Thus,  for  the SN bound RG effects  are somewhat less prominent than in the case of the RGB bound. 
%
For comparison, ignoring the contribution from the pion scatterings,  gives the combination $C_{\rm SN}\simeq 1.5\sqrt{C_{0}^2+0.50\, C_{3}^2 - 0.36\,C_{0}C_{3}}$.
Notice that, as discussed above, the two results have a significantly different dependence on 
$C_{0,3}$ and, in particular, the addition of the pion-induced scatterings increases 
the relative importance of $C_3$ and thus enhances the dependence on the RG corrections. 

\subsection{Thermal axion cosmology}
\label{sec:thermal_axions}

\begin{figure}[t!]
\centering
\includegraphics[width=0.45\textwidth]{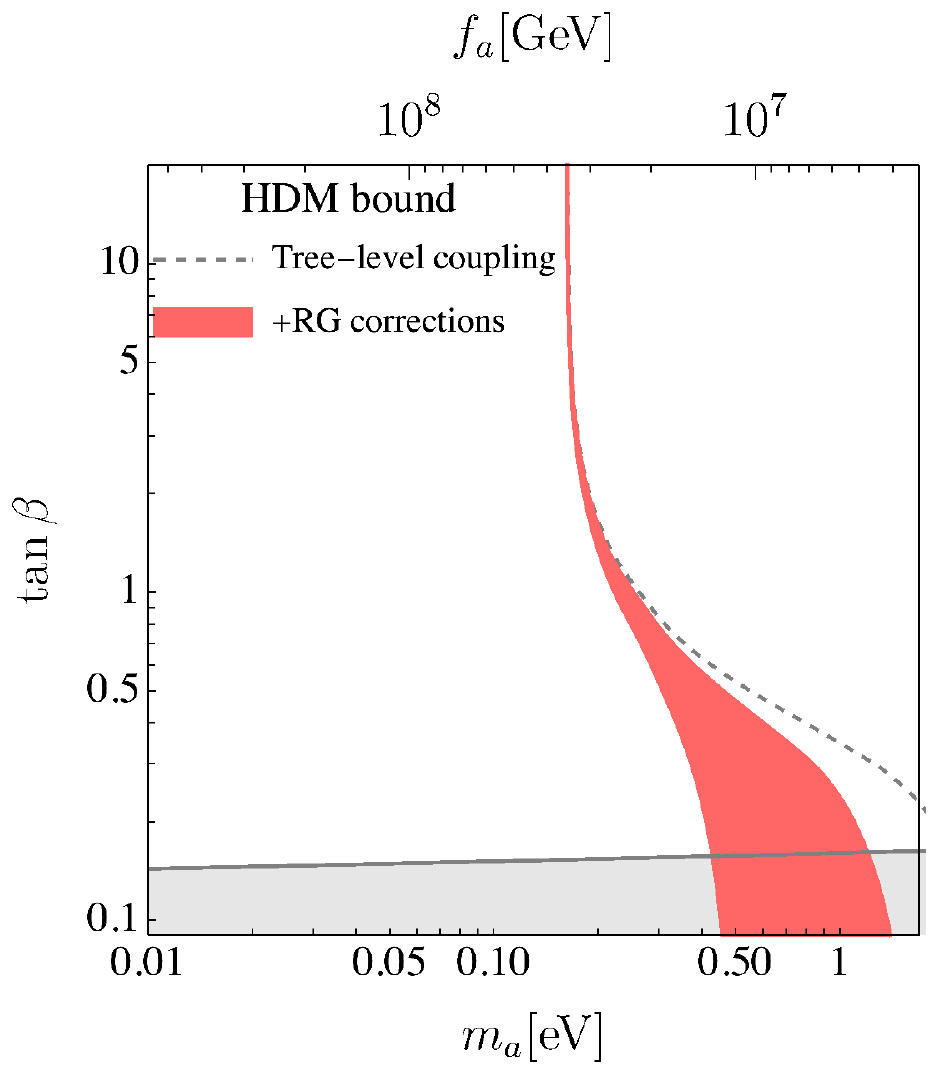}
\caption{HDM bound in the DFSZ1/2 models. 
The red region shows the effect of RG corrections, for $m_{\rm BSM}$ 
ranging from  $f_a$ (left border)
to $1\,$TeV (right border).  
The gray line corresponds to the perturbative 
unitarity bound on $\tan\beta$ 
for $m_{\rm BSM} = f_a$.
\label{fig:HDMbound}}
\end{figure}

If axions are in thermal equilibrium until below the 
quark-hadron phase transition (which can occur for 
$m_a \gsim 0.1\,$eV)  
the axion thermal population will give a sizeable  contribution to the effective number of extra 
relativistic degrees of freedom \cite{Kolb:1990vq}, $\Delta N_{\rm eff}$, that is    
constrained by Big Bang Nucleosynthesis (BBN)~\cite{Cyburt:2015mya} and 
cosmic microwave background (CMB) observations~\cite{Planck:2018nkj,Planck:2018vyg}.
The highest attainable axion mass from such cosmological constraints is also known as the Hot Dark Matter (HDM) bound.
The forecast sensitivity of the planned CMB-S4~\cite{CMB-S4:2016ple} and Simons Observatory (SO)~\cite{SimonsObservatory:2018koc} surveys will fully cover the mass range in which the axion decouples below or across the QCD crossover, thus a precise determination of the axion-pion thermalization rate, including running effects, would be necessary to set definite targets.\footnote{In this paper we will refrain 
from assessing the impact of CMB-S4 and SO projections on $\Delta N_{\rm eff}$, since these 
involve an extrapolation of the axion thermalization rate 
beyond the QCD crossover, which is plagued by large non-perturbative 
uncertainties \cite{Notari:2022zxo}, 
and it is 
still matter of investigation.}

For $T\lesssim T_c$,
where $T_c \simeq 155\,$MeV is the QCD deconfinement temperature, 
the dominant thermalization channels is $a\pi \leftrightarrow \pi\pi$~\cite{Chang:1993gm,Hannestad:2005df}. 
It has been recently shown, however, that the standard 
computation of this process, that is based on 
chiral perturbation theory (ChPT),  breaks down for $T \gtrsim 70\ \rm MeV$ \cite{DiLuzio:2021vjd,DiLuzio:2022gsc}. Phenomenological extensions of the validity of the chiral expansion, based on unitarization methods, 
have been proposed in Refs.~\cite{DiLuzio:2022gsc,Notari:2022zxo}. 

In the following, we will consider the unitarized thermal rate 
based on the Inverse Amplitude Method (IAM), 
recently discussed in Ref.~\cite{DiLuzio:2022gsc}, 
which gives the thermal scattering rate:
\begin{align}
\label{gammaIAM}
  \Gamma_a^{\rm IAM}(T) &= \left(  \frac{C_{\pi}}{f_a f_\pi}\right)^2 
  0.150\ T^5 h_{\rm IAM}(m_\pi/T)  \, ,
\end{align} 
with $C_{\pi}$ given in \eq{eq:defCapi} and 
$m_\pi = 137$\,MeV representing 
the average neutral/charged pion mass. 
The numerical function $h_{\rm IAM}$ is 
provided in Ref.~\cite{DiLuzio:2022gsc} 
(cf.~Fig.~3 of this reference) 
and is normalized to $h_{\rm IAM}(m_\pi/T_c)=1$.

We will estimate the impact of RG effects on the HDM bound 
relying for simplicity on the instantaneous decoupling condition $\Gamma_a(T_D) \simeq H(T_D)$, with  $\Gamma_a(T)$ the axion-pion scattering rate given in \eq{gammaIAM} 
and $ H(T)= \sqrt{4\pi^3 g_\star(T) / 45} \, T^2 / m_{\rm pl}$ the Hubble rate, 
where $m_{\rm pl} = 1.22 \times 10^{19}$\,GeV 
is the Planck mass and $g_\star(T)$  the 
effective number of 
relativistic degrees of freedom.\footnote{For a more 
refined treatment of the cosmological aspects 
of axion thermal decoupling
see e.g.~Refs.~\cite{Caloni:2022uya,DEramo:2022nvb,Notari:2022zxo}.}

The axion contribution to the effective number of extra 
relativistic degrees of freedom is given by
\cite{Kolb:1990vq}
\beq 
\label{eq:DeltaNeff}
\Delta N_{\rm eff} \simeq 
\frac{4}{7} \left(\frac{T_a}{T_\nu}\right)^4 =
\frac{4}{7} 
\( \frac{43}{4 g_S(T_D)} \)^{4/3} 
\simeq 0.027\,
\(\frac{106.75}{g_S(T_D)}\)^{4/3} \, ,  
\eeq
with $T_{a}/T_\nu$ the ratio of the axion to neutrino temperature at $T\ll 1\,$MeV (i.e.~well after 
$\nu$-decoupling) and $g_S(T_D)$ the number of entropy degrees of freedom at  axion decoupling, that in the last relation has been normalised to the total number of SM degrees of freedom $g_S(T>m_t) = 106.75$. 
We then confront \eq{eq:DeltaNeff}
with the   bound on $\Delta N_{\rm eff}$ 
from Planck's 2018 data~\cite{Planck:2018nkj,Planck:2018vyg},
and from this we  extract a bound
on the axion mass. 

Our results for the HDM bound 
in the DFSZ1/2 models are summarised 
in \fig{fig:HDMbound},  
where we show the tree-level results compared 
with the RG corrections included. 
Again we see 
that RG effects  are especially important 
at small $\tan\beta$. 
In Fig.~\ref{fig:HDMbound} the DFSZ1 and DFSZ2 cases coincide because the (subleading) effects of scattering off leptons have been neglected. 
In Ref.~\cite{Ferreira:2020bpb} 
it was argued that thermalization channels 
involving axion scattering off leptons 
can become relevant in DFSZ2 at small $\tan\beta$. 
However, since RG corrections keep the axion-pion coupling sizeable also in this regime,
the effect of lepton scattering becomes accordingly less important.

\subsection{Helioscope experiments}
\label{sec:experiments}

\begin{figure*}[t!]
\centering
\includegraphics[width=0.45\textwidth]{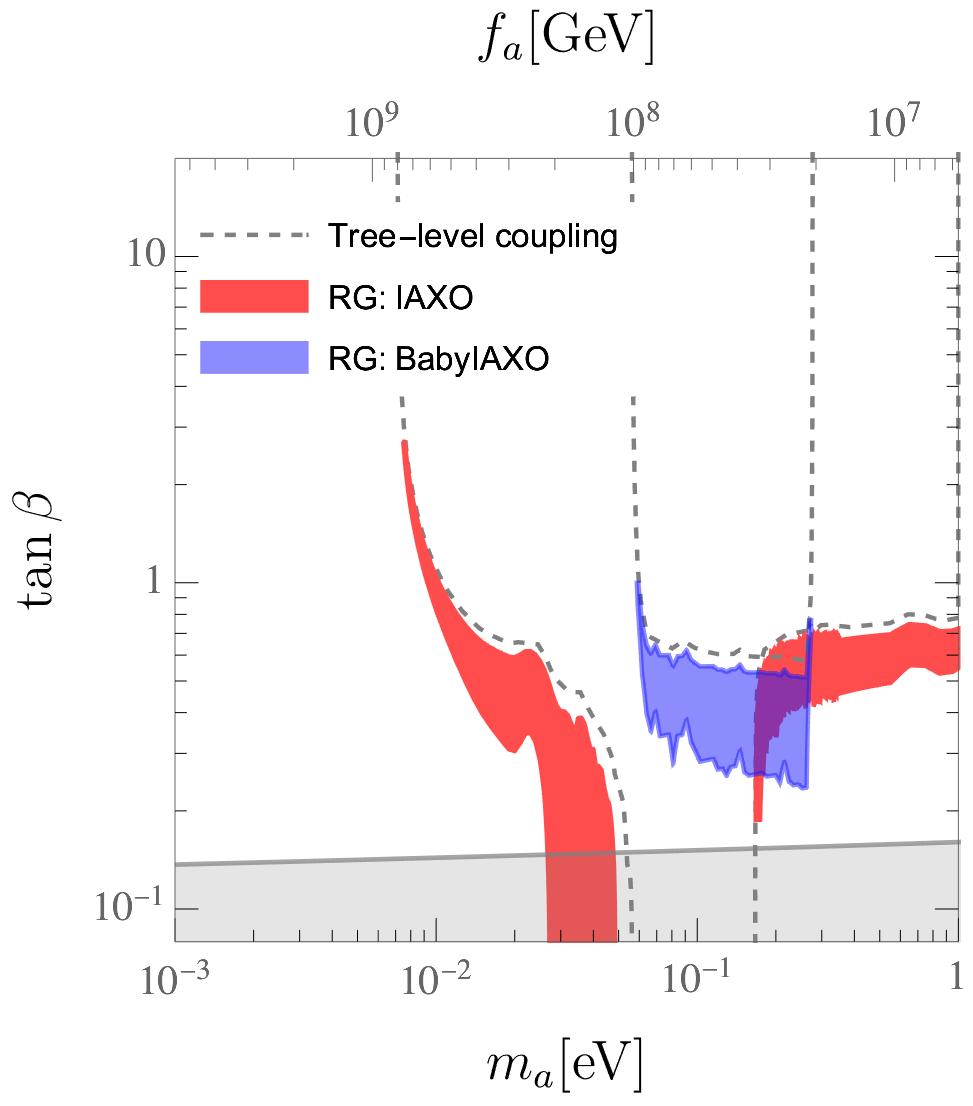}
\qquad
\includegraphics[width=0.45\textwidth]{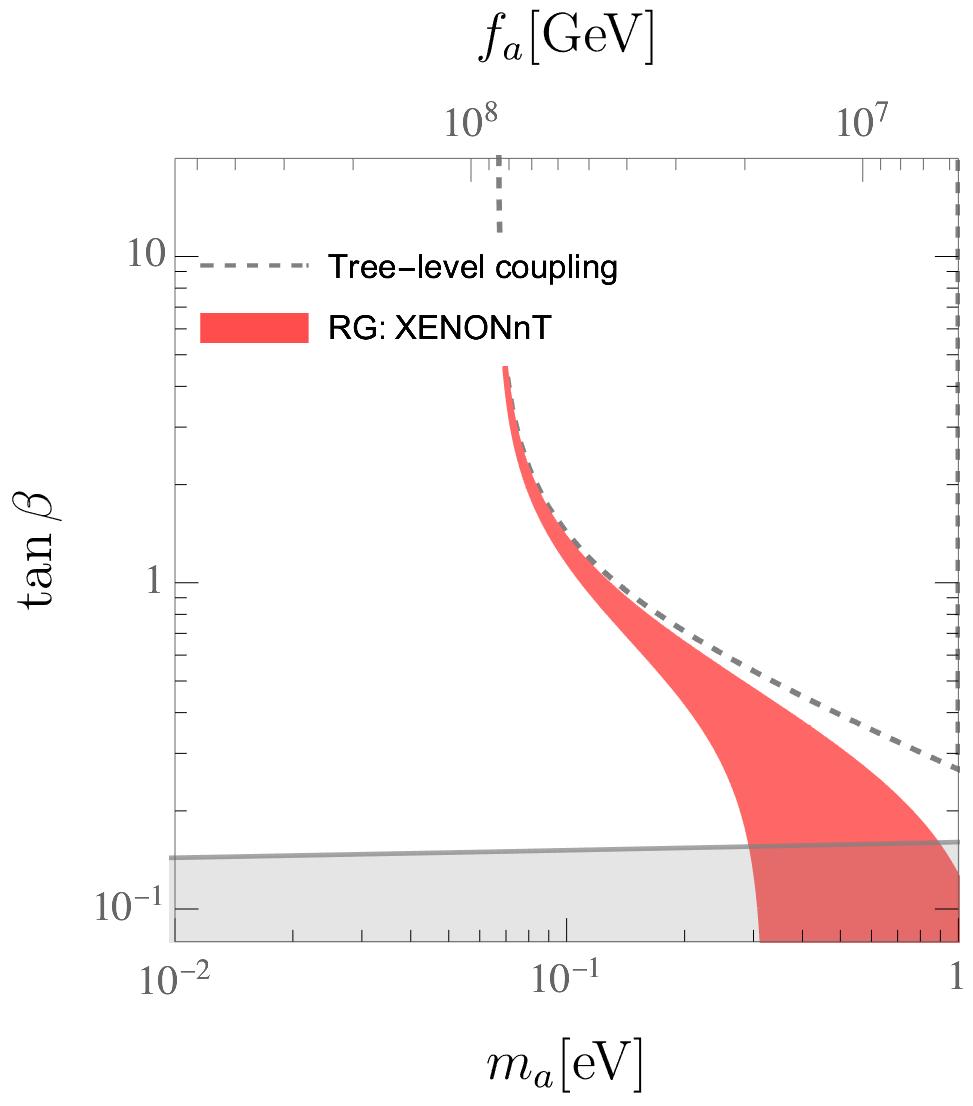}
\caption{Experimental sensitivity to DFSZ1 axions, including RG effects. 
\emph{Left Panel}: IAXO and BabyIAXO.
\emph{Right Panel}: XENON-nT. 
In both plots the gray line corresponds to the perturbative 
unitarity bound on $\tan\beta$ 
for $m_{\rm BSM} = f_a$.
\label{fig:DFSZ1_IAXO}}
\end{figure*}

One of the most appealing result of the RG correction analysis is the implication for the next generation of experiments hunting for solar axions. 
The main reason is that the solar flux is strongly dependent on the axion-electron coupling and, as we have seen, this can receive large RG corrections.
As a consequence, helioscope sensitivities to DFSZ axions, that have been so far estimated using  
tree level electron-axion couplings, have been
underestimated. 

Here, we focus mostly on the Sikivie type of axion helioscopes~\cite{Sikivie:1983ip}.
This kind of experiment is designed to detect solar axions by converting them in X-ray photons using a large laboratory magnetic field. 
The importance of the axion-electron coupling for Sikivie's helioscope sensitivity to solar axions is expressed by the following relation~\cite{DiLuzio:2020wdo} 
\begin{align}
\label{eq:gag-heli}
g_{\gamma 10}^{2}\left( g_{\gamma 10}^{2} +0.7 g_{e 12}^{2}\right) > \bar{g}_{\gamma 10}^{4}\,,
\end{align}
where $g_{\gamma 10}=g_{a\gamma}/10^{-10}{\rm GeV}^{-1}$, $g_{e12}=g_{ae}/10^{-12}\,$, and $\bar{g}_{\gamma 10}$ is the helioscope sensitivity to $g_{a\gamma}$ (again, in units of $10^{-10}{\rm GeV}^{-1}$).
Notice that $\bar{g}_{a\gamma}$ is, in general, a function of the axion mass.
Defining the effective coupling 
\begin{align}
\label{eq:C_helioscope}
C_{\rm hel}=\left[ C_{\gamma}^2\left( C_{\gamma}^2 +(37 C_e)^2 \right) \right]^{1/4}\,,
\end{align}
the above expression leads to the following heliscope sensitivity relation
\begin{align}
\label{eq:helioscope_sensitivity}
C_{\rm hel}\gtrsim \frac{0.49 \,\bar{g}_{\gamma 10}}{(m_a/{\rm eV})}\,,
\end{align}
which, in the case of the DFSZ axion, can be readily translated into a limit on the $\tan\beta$ accessible to the helioscope as a function of the axion mass. 
Notice that, according to this expression, the DFSZ sensitivity to $\tan\beta$ (which enters only through $C_e$), should disappear for $C_\gamma\gg 37 C_e$, which is fulfilled for $\tan\beta\ll 0.25$.
In general, if the helioscope sensitivity is good enough, there could be mass regions where the entire range of $\tan \beta$ is accessible.

To give an example of an application of 
Eq.~\eqref{eq:helioscope_sensitivity}, let us consider the case of BabyIAXO, a next-generation axion helioscope presently under construction~\cite{IAXO:2020wwp}. 
Its sensitivity at $m_a=0.1\,$eV is expected to reach $\bar{g}_{\gamma 10}=0.33$.
Using this value and $C_\gamma=8/3-1.92$ for the DFSZ1 model, if RG effects are ignored, one would conclude that at this mass value BabyIAXO could be sensitive to the region $\tan\beta\gtrsim 0.62$.
The results of our complete numerical analysis for all values of the axion mass are plotted 
in the left panel in Fig.~\ref{fig:DFSZ1_IAXO}, 
where the  dashed line contours 
correspond  to the estimated BabyIAXO sensitivity if RG effects are ignored.
The reach of the more advanced helioscope experiment IAXO~\cite{IAXO:2019mpb} is also shown in the left panel in Fig.~\ref{fig:DFSZ1_IAXO}.
In this case we see that there is a mass region for which the experiment is sensitive to all values of $\tan\beta$.
When RG corrections are ignored, this region extends to masses between $\sim 50$ meV and $\sim 200$ meV.
The reason of this is that IAXO is sensitive enough to see the solar axion flux even in models in which axions are only coupled to the photon and not to the electron. 

Let us now consider the effects of RG corrections on the projected sensitivities. 
As shown in Tab.~\ref{tab:RG_corrections}, the RG corrections to the effective helioscope coupling is
\begin{align}
\label{eq:delta_C_hel}
\frac{\Delta C_{\rm hel}}{C_{\rm hel}}
	=\dfrac{1}{2} \dfrac{\Delta C_e/C_e}{1+\left({C_\gamma}/(37C_e)\right)^2} \,,
\end{align}
which is valid in the limit of $\Delta C_e/C_e \ll 1$.
This condition is always verified in  DFSZ2, while 
for DFSZ1 it holds for $\tan\beta\gg 0.5\,l(x)^{1/2}$ (Cf. Tab.~\ref{tab:DFSZ_corrections}).
Since in the case of BabyIAXO the expected sensitivity is not sufficient to detect DFSZ axions unless $\tan\beta\gtrsim 0.6$ (see  Fig.~\ref{fig:DFSZ1_IAXO}) which implies $C_\gamma/(37 C_e)\ll 1$, 
we can simplify the correction to the effective coupling to 
\begin{align}
\left(\frac{\Delta C_{\rm hel}}{C_{\rm hel}}\right)^{\rm BabyIAXO}\simeq 
\frac{\Delta C_e}{2C_e}\,, 
\end{align}
which can be readily estimated using our results from the Tab~\ref{tab:DFSZ_corrections}. 
Notice that this correction can be quite sizable and implies that the reach of BabyIAXO to small electron couplings (low $\tan\beta$ values) can be pushed down significantly, as is shown by the blue region in the left panel of Fig.~\ref{fig:DFSZ1_IAXO}.

The impact of RG corrections  on the IAXO sensitivity to DFSZ1 axions is also shown in the left panel of Fig.~\ref{fig:DFSZ1_IAXO}, and corresponds to the red regions. 
In this case we notice an interesting effect, that is that the 
 IAXO reach in the region of small $\tan\beta$ is sizeably enlarged 
for all values of $m_{\rm BSM}$, since  
the solar axion flux is necessarily larger than what predicted ignoring the RG corrections. 
As a result, the mass region for which IAXO is sensitive to the entire range of $\tan\beta$ is extended. 

Finally, the correction to the axion-electron coupling has also an obvious impact on experiments 
which detect axions through the axio-electric effect. 
Such experiments include large underground detectors such as Panda-X~\cite{PandaX-II:2020udv},
LUX~\cite{LUX:2017glr}, or
XENON-nT~\cite{XENON:2022ltv}, originally designed for dark matter searches.
The RG modification of the axion-electron coupling extends the potential of these experiments to explore the DFSZ parameter space. 
Our numerical results in the case of XENON-nT are shown in the right panel of Fig.~\ref{fig:DFSZ1_IAXO}.
A fundamental difference with respect to the previous results is that, because of the RG-induced corrections, 
in principle XENON-nT could be sensitive to DFSZ1 axions for any value of $\tan\beta$.
However, the current experimental sensitivity is insufficient to reach inside the mass region  allowed by the RGB bound discussed in Sect.~\ref{sec:astrophysics}
(see the left panel in Fig.~\ref{fig:DFSZ_astro}).


\section{Conclusions}
\label{sec:concl}

In this paper, we have studied the impact of RG effects on QCD axion phenomenology, focusing on DFSZ models.  
We have shown that running effects on axion couplings depend crucially on the scale at which the heavy Higgs states are integrated out, and 
the 2HDM effectively reduces to the SM with a single light Higgs.
We have discussed the implications of running axion couplings on astrophysical and cosmological bounds, as well as the sensitivity of helioscope experiments such as (Baby)IAXO. 
We have found that running effects are sizable even in the most conservative case in which the 2HDM structure keeps holding down to the TeV-scale, and 
thus they can never be neglected.
We have  also provided simple analytic expressions fitted to reproduce the numerical solutions of the RG equations, which can  be a useful tool for studying the implications of running axion couplings.  
In the case of an axion discovery, 
running effects might prove to be  
crucial in order to reconstruct the axion UV completion.

\section*{Acknowledgments}

We thank Kiwoon Choi and Giovanni Villadoro for useful discussions. 
The work of L.D.L. is funded by the European Union -- NextGenerationEU and by the University of Padua under the 2021 STARS Grants@Unipd programme (Acronym and title of the project: CPV-Axion -- Discovering the CP-violating axion) and by the INFN Iniziative Specifica APINE.
The work of L.D.L. and G.P.~is supported by the European Union's Horizon 2020 research and innovation programme under the Marie Sk\l{}odowska-Curie grant agreement No 860881-HIDDEN.
The work of E.N.~was supported  by the Estonian Research Council grant
PRG1884 and by the INFN ``Iniziativa Specifica" Theoretical Astroparticle Physics (TAsP-LNF). 
S.O.~and F.M.~acknowledge financial support from a Maria Zambrano fellowship and the State Agency for Research of the Spanish Ministry of Science and Innovation
through the Unit of Excellence Mar\'ia de Maeztu 2020-2023 award to the Institute of Cosmos Sciences (CEX2019-000918-M)
and from PID2019-105614GB-C21, 2017-SGR-929 and 2021-SGR-249 grants. This article is based upon work from COST Action COSMIC WISPers CA21106,
supported by COST (European Cooperation in Science and Technology).
M.G. research was partially supported by funds from 
``European Union NextGenerationEU/PRTR'' 
(Planes complementarios, Programa de Astrof\'isica y F\'isica de Altas Energías) and 
from the European Union's Horizon 2020 research and 
innovation programme under the European Research Council (ERC) grant agreement ERC-2017-AdG788781 (IAXO+).
We thank the Galileo Galilei Institute for Theoretical Physics for hospitality during the completion of this work.

\appendix

\section{RG equations for DFSZ axion couplings}
\label{sec:AxionEFT}

In order to take into account running effects
it is convenient to adopt the Georgi-Kaplan-Randall (GKR) field basis \cite{Georgi:1986df}, 
where the PQ symmetry is realised non-linearly, so 
that under a $\U(1)_{\rm PQ}$ symmetry transformation 
all fields are invariant except for the axion field, 
which changes by an additive constant 
$a \to a + \alpha f$, that is 
\begin{align} 
\label{eq:LaGKRH12}
&\mathcal{L}^{\rm GKR-2HDM}_a = \frac{1}{2} \partial_\mu a \partial^\mu a 
+ \sum_{A=G,W,B} c_A \frac{g_A^2}{32\pi^2} \frac{a}{f} F^A \tilde F^A \\
&\quad + \frac{\partial_\mu a}{f} 
\Big[ c_{H_u} H_u^\dag i \overleftrightarrow{D^\mu} H_u + c_{H_d} H_d^\dag i \overleftrightarrow{D^\mu} H_d 
+ \bar q_L c_{q_L} \gamma^\mu q_L  \nonumber \\
&\quad + \bar u_R c_{u_R} \gamma^\mu u_R 
+ \bar d_R c_{d_R} \gamma^\mu d_R
+ \bar \ell_L c_{\ell_L} \gamma^\mu \ell_L 
+ \bar e_R c_{e_R} \gamma^\mu e_R 
\Big] \, , \nonumber
\end{align}
where $H_{u,d}^\dag \overleftrightarrow{D^\mu} H_{u,d} \equiv H_{u,d}^\dag (D^\mu H_{u,d}) - (D^\mu H_{u,d})^\dag H_{u,d}$
and $c_{q_L}, \ldots $ are diagonal matrices in generation space. 
Note that in the effective field theory below 
$f$ we have neglected the heavy $\mathcal{O}(f)$ radial mode of $\Phi$   
and we focused for simplicity on the 2HDM.   
In order to match an explicit axion model 
to the effective Lagrangian in \eq{eq:LaGKRH12} at the high scale $\mu \sim \mathcal{O}(f)$, 
we perform an axion dependent field redenfinition: 
$\psi \to e^{-i \X_\psi a /f} \psi$, where $\psi$ spans over all the fields, and $\X_\psi$ is the corresponding PQ charge. 
Due to $\U(1)_{\rm PQ}$ symmetry, 
the non-derivative part of the renormalizable Lagrangian is 
 invariant upon this field redefinition, 
while the $d=5$ operators in 
\eq{eq:LaGKRH12} 
are generated 
from the variation of the kinetic terms and from the chiral anomaly. The couplings are then identified as  
\begin{align} 
\label{eq:cpsi}
c_{\psi} &= \X_\psi \, , \\
\label{eq:cA}
c_A & =
\sum_{\psi_R} 2 \X_{\psi_R} \Tr T^2_A(\psi_R) -
\sum_{\psi_L} 2 \X_{\psi_L} \Tr T^2_A(\psi_L)  
\, , 
\end{align}
where in the second equation 
$c_{\psi_{R,L}}$ refer to the charges of the chiral fermion fields.\footnote{Note that our anomaly coefficients $c_A$ have opposite sign with respect to those
in Refs.~\cite{Choi:2017gpf,Bauer:2020jbp,Choi:2021kuy}. This is due to the
fact that we are using a different convention for the Levi-Civita tensor, namely $\epsilon^{0123}=-1$.}
For the DFSZ1 model    
introduced in \sect{sec:DFSZ_axion_couplings}, 
with e.g.~the operator $H_u H_d \Phi^{\dag}$ 
in the scalar potential,
the universal charges $\X_\psi$  
can be set to
\begin{align}
\label{eq:M1charges}
\X_{q} = \X_{\ell} = 0\, ,\
\X_{u} = - \X_{H_u}\, ,\ 
\X_{d} = - \X_{H_d} \, , \ 
\X_{e}= - \X_{H_d} 
\, ,
\end{align}
where $\X_{H_u} = c^2_{\beta}$ and 
$\X_{H_d} = s^2_{\beta}$. 
The DFSZ2 model features a similar charge assignment, 
with instead $\X_{e}= \X_{H_u}$. 
For the anomaly coefficients in \eq{eq:cA} in the case of the DFSZ1 model 
one has $(c_G, c_W, c_B) = (-3, 0, -8)$ 
and, in particular, the electromagnetic to QCD 
anomaly ratio is $E/N \equiv (c_W + c_B) / c_G = 8/3$. 
Similarly, for the DFSZ2 model one finds 
$(c_G, c_W, c_B) = (-3, 0, -2)$ and $E/N = 2/3$. 

Running effects induced by Yukawa couplings (and in particular by the top Yukawa 
which is the most relevant one) 
only occur below the scale of the heavy radial modes of the 2HDM, 
denoted as 
$m_{\rm BSM} \simeq m_{H,\,A,\,H^\pm}$,  with the heavy scalars  assumed to be degenerate in the decoupling limit (see e.g.~\cite{Gunion:2002zf}). 
This is due to the fact that as long as the complete set  of Higgs doublets  appear in the effective field theory, 
the PQ current is conserved (up to anomalous effects) and thus the couplings, which correspond to PQ charges, do not renormalize. Once the heavy scalar components are 
integrated out,  the sum-rule of PQ charges set by $\U(1)_{\rm PQ}$ invariance  breaks down,  and non-vanishing contributions to 
the running of the couplings arise (see e.g.~\cite{Choi:2021kuy}). 
We can now directly match \eq{eq:LaGKRH12} 
at the scale $\mu = \mathcal{O}(m_{\rm BSM})$
with a GKR  basis  featuring  only one SM-like Higgs doublet 
\begin{align} 
\label{eq:LaGKRH}
\mathcal{L}^{\rm GKR-SM}_a &= \frac{1}{2} \partial_\mu a \partial^\mu a 
+ \sum_{A=G,W,B} c_A \frac{g_A^2}{32\pi^2} \frac{a}{f} F^A \tilde F^A \\
&+ \frac{\partial_\mu a}{f} 
\Big[ c_{H} H^\dag i \overleftrightarrow{D^\mu} H  
+ \bar q_L c_{q_L} \gamma^\mu q_L  \nonumber \\
&+ \bar u_R c_{u_R} \gamma^\mu u_R 
+ \bar d_R c_{d_R} \gamma^\mu d_R
+ \bar \ell_L c_{\ell_L} \gamma^\mu \ell_L 
+ \bar e_R c_{e_R} \gamma^\mu e_R 
\Big] \, , \nonumber
\end{align}
where 
$c_{H} = c_{H_u} s^2_\beta - c_{H_d} c^2_\beta$,  
which follows from the projections 
on the SM Higgs doublet $H\sim(1,2,-1/2)$: 
$H_u \to s_\beta \, H$ and  
$H_d \to c_\beta \, \tilde{H}$,  
consistently with the definition 
of $\tan\beta = v_u/v_d$. 
In particular, by employing global $\U(1)_Y$ invariance, 
it is convenient to cast 
the RG equations in a 
form that does not depend explicitly on $c_H$. 
This can be achieved via  
the axion-dependent field redefinition:  
$\psi \to \psi' = e^{-i c_H \beta_\psi a/f} \psi$,  
with $\beta_\psi = Y_\psi / Y_H$
the ratio of the corresponding hypercharges, 
which redefines the effective couplings as $c'_\psi = c_\psi - c_H \beta_\psi$ 
(so in particular $c'_H = 0$). 
In this basis the RG equations read: 
 \begin{align}
\label{eq:cazzoU1Y}
(4\pi)^2 \frac{dc'_\qL}{d\log\mu} 
           & = \frac{1}{2} \{ c'_\qL, Y_u Y_u^\dagger + Y_d Y_d^\dagger \} 
                   - Y_u c'_\uR Y_u^\dagger - Y_d c'_\dR Y_d^\dagger \nonumber\\
           &+ \left( 8 \alpha_s^2 \widetilde{c}_G + \frac{9}{2} \alpha_2^2 \widetilde{c}_W 
                  + \frac{1}{6} \alpha_1^2 \widetilde{c}_B \right) \, \1  -\b_q \, 
\gamma_H \, \1
\, , \nonumber\\
(4\pi)^2 \frac{dc'_\uR}{d\log\mu} 
           & = \{ c'_\uR, Y_u^\dagger Y_u \} - 2 Y_u^\dagger c'_\qL Y_u 
                  - \left( 8 \alpha_s^2 \widetilde{c}_G + \frac{8}{3} \alpha_1^2 \widetilde{c}_B \right) \, \1 \nonumber \\ 
                  &-\b_u \, 
\gamma_H \, \1
                  \, , \nonumber\\
(4\pi)^2 \frac{dc'_\dR}{d\log\mu} 
           & = \{ c'_\dR, Y_d^\dagger Y_d \} - 2 Y_d^\dagger c'_\qL Y_d 
                  - \left( 8 \alpha_s^2 \widetilde{c}_G + \frac{2}{3} \alpha_1^2 \widetilde{c}_B \right) \, \1 \nonumber \\ 
                  &-\b_d \, 
\gamma_H \, \1
\, , \nonumber\\
(4\pi)^2 \frac{dc'_\lL}{d\log\mu} 
          & = \frac{1}{2} \{ c'_\lL, Y_e Y_e^\dagger \} - Y_e c'_\eR Y_e^\dagger  
                  + \left( \frac{9}{2} \alpha_2^2 \widetilde{c}_W + \frac{3}{2} \alpha_1^2 \widetilde{c}_B \right) \, \1 \nonumber \\ 
                  &-\b_\ell \, 
\gamma_H \, \1
\, , \nonumber \\
(4\pi)^2 \frac{dc'_\eR}{d\log\mu} 
          & = \{ c'_\eR, Y_e^\dagger Y_e \} - 2 Y_e^\dagger c'_\lL Y_e  
                  - 6 \alpha_1^2 \widetilde{c}_B \, \1-\b_e \, 
\gamma_H \, \1
\, , 
\end{align}
where 
\begin{align}
\label{eq:Xdef}
\gamma_H
           & =
                   - 2 \, \tr\( 3Y_u^\dagger c'_\qL Y_u - 3Y_d^\dagger c'_\qL Y_d - Y_e^\dagger c'_\lL Y_e \) \nonumber \\
&+ 2 \, \tr\( 3Y_u c'_\uR Y_u^\dagger - 3Y_d c'_\dR Y_d^\dagger - Y_e c'_\eR Y_e^\dagger \) \, , \nonumber \\
\widetilde{c}_G
           & = c_G - \tr\( c'_{u_R} + c'_{d_R}- 2c'_{q_L}  \) \, , \nonumber\\
\widetilde{c}_W
           & = c_W + \tr\( 3c'_{q_L} + c'_{\ell_L} \) \, , \nonumber\\
\widetilde{c}_B
           & = c_B - \tr\( \frac{1}{3} ( 8c'_{u_R} + 2c'_{d_R} -c'_{q_L} ) + 
           2 c'_{e_R}- c'_{\ell_L} \) \, .
\end{align}
Note that the $c_A$ ($A = G,W,B$) Wilson coefficients in \eq{eq:Xdef} do not run at one loop,   
since in the normalization of \eq{eq:LaGKRH12} the scale dependence of the operator $a F^A\tilde F^A$  
is  accounted for by the running of the gauge 
couplings \cite{Bauer:2020jbp,Chetyrkin:1998mw}.  

\eq{eq:LaGKRH} is matched at the scale $\mu = \mathcal{O}(m_Z)$ with the 
$\SU(3)_C \times \U(1)_{\rm EM}$-invariant
axion effective Lagrangian below the electroweak 
scale
\begin{align}
\label{eq:LeffEW}
\mathcal{L}_a &\supset \mathcal{L}_{GF} 
+ \mathcal{L}_f \, , \\
\mathcal{L}_{GF}&=\frac{g_s^2}{32\pi^2} \frac{a}{f_a} G \tilde G + 
\frac{c_\gamma}{c_G} \frac{e^2}{32\pi^2} \frac{a}{f_a} F \tilde F \, , 
\\
\mathcal{L}_{f}&=\sum_{f = u,\,d,\,e} \frac{\partial_\mu a}{2 f_a} \bar f_i \gamma^\mu \left[  (C^V_f)_{ij} + (C^A_f)_{ij} \gamma_5 \right] f_j \, , 
\end{align}
where we have introduced the standard QCD  normalization 
factor for the $aG\tilde G$ term and defined 
the axion decay constant $f_a = f / c_G$, while 
$c_\gamma = c_W + c_B$. We further have  
\begin{align}
\label{eq:CfV}
C^V_f &= \frac{1}{c_G} ( U_{f_R} c'_{f_R} U_{f_R}^\dag + U_{f_L} c'_{f_L} U_{f_L}^\dag ) \, , \\
\label{eq:CfA}
C^A_f &= \frac{1}{c_G} ( U_{f_R} c'_{f_R} U_{f_R}^\dag - U_{f_L} c'_{f_L} U_{f_L}^\dag ) \, , 
\end{align}
where $U_{f_{L,R}}$ are  the unitary matrices that diagonalize the SM fermion mass matrices, and 
$c'_{u_L} = c'_{d_L}  = c'_{q_L}$.  
 Note that the diagonal vector couplings $(C^{V}_{f})_{ii}$ can be 
always set to zero thanks to the conservation of the vector current. 
In our case, 
since the models that  we have considered enjoy flavour universality, the matrices of couplings $c'_{f_{R,L}}$ are proportional to the identity, then $\mathcal{L}_f$ simplifies to: 
\begin{align}
\label{eq:Cfuniv}
\mathcal{L}_{f} &= \sum_{f = u,\,d,\,e} \, C^A_f\, \frac{\partial_\mu a}{2 f_a} \,\bar f \gamma^\mu \gamma_5 f\, , \\
C^A_f &= \frac{1}{c_G} ( c'_{f_R} 
- c'_{f_L}  ) \, .
\end{align}
After including matching corrections 
at the weak scale \cite{Bauer:2020jbp}, 
the running for $\mu < m_Z$ is given by 
\begin{align}
\label{eq:rges}
(4\pi)^2 \frac{dC^A_{u}}{d\log\mu} 
           & = - 16 \alpha_s^2 \wt{c}_G - \frac{8}{3} \alpha_{\rm em}^2 \wt{c}_\gamma \, , \nonumber\\
(4\pi)^2 \frac{dC^A_{d}}{d\log\mu} 
           & = - 16 \alpha_s^2 \wt{c}_G -  \frac{2}{3} \alpha_{\rm em}^2 \wt{c}_\gamma \, , \nonumber \\
(4\pi)^2 \frac{dC^A_{e}}{d\log\mu} 
           & = - 6 \alpha_{\rm em}^2 \wt{c}_\gamma \, , 
\end{align}
with 
\begin{align}
\label{eq:ctildeG}
\wt{c}_G(\mu) &= 1 - \sum_q C^A_q(\mu) \Theta(\mu-m_q) \, , \\
\label{eq:ctildegamma}
\wt{c}_\gamma(\mu) &= \frac{c_\gamma}{c_G} - 2 \sum_f N_c^f Q_f^2 C^A_f(\mu) \Theta(\mu-m_f) \, ,
\end{align}
where $\Theta(x)$ is the Heaviside theta function, while 
$N_c^f$ and $Q_f$ denote respectively the colour number and EM charge of the fermion $f$. 

The axion-nucleon couplings, neglecting the tiny contributions of the matrix elements 
$\Delta_{t,b,c}$ of the heavy flavours, can be calculated by using \eqs{eq:CpDelta}{eq:CnDelta},  
with $C_{u,d,s} = C^A_{u,d,s}(2\GeV)$ 
evaluated by numerically solving the RG equations, Eqs.~(\ref{eq:cazzoU1Y}) and (\ref{eq:rges}), 
starting from the boundary conditions set at the scale $f$ 
(cf.~below \eq{eq:cA}).

\section{Numerical fit to RG effects}
\label{sec:RGEfit}

Running of axion couplings is examined in detail in Refs.~\cite{Choi:2017gpf,Bauer:2020jbp,Choi:2021kuy,Choi:2020rgn}, 
where a complete set of one-loop (and partially two-loop) anomalous dimensions are derived 
including matching corrections at the EW scale~\cite{Bauer:2020jbp}.
The leading contribution to the running axion couplings arises from 
top loop diagrams induced by the axion-top coupling $C_t$.
The RG evolved couplings at $\mu=2\GeV$ are thus expressed 
to a good approximation by 
\beq
C_\Psi (2\GeV) \simeq C_\Psi(f_a) + r_\Psi^t (m_\BSM) \, C_t(f_a),
\label{eq:CPsi}
\eeq
where $\Psi=u,d,e$.
Note that the running occurs below the heavy Higgs scale $m_\BSM \simeq m_{H,A,H^+}$, 
where in the decoupling limit the heavy scalars are assumed to be approximately degenerate, 
and $r_\Psi^t(m_\BSM)$ is a function only of $m_\BSM$. 

Keeping only the top Yukawa and the strong gauge couplings,
the running of $C_\Psi$ below $\mu=m_\BSM$ is governed by \cite{Choi:2017gpf,Bauer:2020jbp,Choi:2021kuy,Choi:2020rgn}
\beq
\frac{d C_\Psi}{d \ln\mu} 
	\simeq - T_{3,\Psi} \frac{3Y_t^2}{4\pi^2} \, C_t \, \Theta(\mu-\mu_w)
		- a_\Psi \frac{\alpha_s^2}{\pi^2} \wt{c_G} \,,
\label{eq:RGEs}
\eeq
where $T_{3,\Psi}$ is the weak isospin of $\Psi$, 
$a_\Psi = 1$ for quarks and $0$ for leptons, 
$\mu_w={\cal O}(m_Z)$ is a matching scale at which 
weak gauge bosons,  Higgs boson and  top quark
are integrated out, 
with $\wt{c}_G$ defined in \eq{eq:ctildeG}.

We see from Eq.~(\ref{eq:RGEs}) that 
the RG corrections to the axion couplings consist of 
one-loop iso-vector contribution, 
proportional to the weak isospin $T_{3,\Psi}$, and a
two-loop level iso-scalar contribution 
generated from $\wt{c}_G$,\footnote{In the DFSZ models, 
$\wt{c}_G=0$ at $\mu=m_\BSM$ and 
it develops a nonzero value because of the running of $C_q$.
This means that 
the running effects from $\wt{c}_G$ 
are also proportional to $C_t(f_a)$, 
allowing to parametrise this iso-scalar contribution 
in the form of \eq{eq:CPsi}.} 
and can be expressed in the form
\beq
r_\Psi^t(m_\BSM) \simeq T_{3,\Psi} \, r^t_3 (m_\BSM) + \frac{a_\Psi}{2} \, r^t_0 (m_\BSM) \,,
\eeq
which, for the running of $C_{3,0}$,  yields
\begin{align}
\label{eq:rt3}
r^t_3 & \simeq r^t_u -  r^t_d \simeq -2r^t_e \,, \\
\label{eq:rt0}
r^t_0 & \simeq r^t_u +  r^t_d \,.
\end{align}
Note that 
$r^t_{3,0}$ are independent of $\Psi$ 
to a good precision,  
even after including the threshold corrections at the EW scale, 
which turn out to be iso-vector 
(numerically $\left|(r^t_0)_{\rm th}/(r^t_3)_{\rm th}\right| \sim 10^{-6}$). 

Let us now derive approximate formulae for $r^t_{3,0} (m_\BSM)$. 
To this end, 
we first evaluate the running effects 
by numerically solving the full set of the RG 
equations  
including the threshold corrections at the EW scale~\cite{Bauer:2020jbp}. 
In the calculation 
the two-loop running for the SM gauge and Yukawa couplings is implemented, with  
their input values at $\mu_w=m_Z$  taken from Ref.~\cite{Antusch:2013jca}. 
A set of numerical values for $r^t_{3,0} (m_\BSM)$ are tabulated 
in Tab.~\ref{tab:rpsit}. These values are 
accurately fitted by the following fitting functions:
\begin{align}
\label{eq:rt3fit}
r^t_3(m_\BSM) & \simeq  r_u^t-r_d^t \simeq -0.54 \ln \left( \sqrt{x} - 0.52 \right) \,,\\
\label{eq:rt0fit}
r^t_0(m_\BSM) & \simeq  r_u^t+r_d^t \simeq 3.8\times 10^{-4} \ln^2 \left( x - 1.25 \right) \,,
\end{align}
with $x=\log_{10}(m_\BSM/\GeV)$. 
\eq{eq:rt3fit} agrees with the numerical results within 2\% accuracy in the $1\TeV \leq m_\BSM \leq 10^{18} \GeV$ range.
The precision of \eq{eq:rt0fit} is better than $6\%$. 
However,  since $|r^t_0/r^t_3| \lesssim 0.5\%$, 
this function does not affect numerically $r^t_\Psi$.

\begin{table*}[!ht]
\renewcommand{\arraystretch}{1.1}
\centering
\begin{tabular}{|c|c|c|c||c|c|}
\hline
 $m_{\rm BSM}$ [GeV] &  $r_u^t$  &  $r_d^t$  &  $r_e^t$  &  $r^t_3=r^t_u-r^t_d$  &  $r^t_0=r^t_u+r^t_d$  \\[0.6ex]
 \hline 
$10^3$  & ~~$-0.0523595$~~ & ~~$0.0524821$~~ & ~~$0.0524214$~~ & ~~$-0.104842$~~ & ~~$0.000122546$~~ \\
$10^4$  & $-0.104883$ & $0.105251$ & $0.105072$ & $-0.210134$ & $0.000368801$ \\
$10^5$  & $-0.145433$ & $0.146074$ & $0.145764$ & $-0.291507$ & $0.000640706$ \\
$10^6$  & $-0.177998$ & $0.178907$ & $0.178469$ & $-0.356906$ & $0.000909250$ \\
$10^7$  & $-0.204893$ & $0.206057$ & $0.205498$ & $-0.410949$ & $0.00116422$  \\
$10^8$  & $-0.227574$ & $0.228976$ & $0.228305$ & $-0.456550$ & $0.00140241$  \\
$10^9$  & $-0.247016$ & $0.248639$ & $0.247865$ & $-0.495655$ & $0.00162345$  \\
$10^{10}$  & $-0.263900$ & $0.265729$ & $0.264859$ & $-0.529629$ & $0.00182809$ \\
$10^{11}$  & $-0.278732$ & $0.280749$ & $0.279793$ & $-0.559481$ & $0.00201769$ \\
$10^{12}$  & $-0.291859$ & $0.294052$ & $0.293015$ & $-0.585911$ & $0.00219320$ \\
$10^{13}$  & $-0.303574$ & $0.305930$ & $0.304819$ & $-0.609504$ & $0.00235605$ \\
$10^{14}$  & $-0.314096$ & $0.316603$ & $0.315424$ & $-0.630699$ & $0.00250738$ \\
$10^{15}$  & $-0.323599$ & $0.326247$ & $0.325006$ & $-0.649847$ & $0.00264829$ \\
$10^{16}$  & $-0.332225$ & $0.335005$ & $0.333705$ & $-0.667230$ & $0.00277971$ \\
$10^{17}$  & $-0.340088$ & $0.342991$ & $0.341637$ & $-0.683079$ & $0.00290251$ \\
$10^{18}$  & $-0.347283$ & $0.350300$ & $0.348897$ & $-0.697583$ & $0.00301745$ \\
\hline
\end{tabular}
    \caption{ 
    The numerical values of 
    $r_\Psi^t(m_{\rm BSM})$ and $r^t_{3,0}(m_{\rm BSM})$, 
    which are obtained by numerically solving the full RGEs, with 
     the threshold corrections at the EW scale included. 
    \label{tab:rpsit} }
\end{table*}


\bibliographystyle{utphys}
\bibliography{bibliography}

\providecommand{\href}[2]{#2}\begingroup\raggedright\begin{thebibliography}{10}

\bibitem{Peccei:1977hh}
R.~D. Peccei and H.~R. Quinn, ``{CP Conservation in the Presence of
  Instantons},''
\href{http://dx.doi.org/10.1103/PhysRevLett.38.1440}{{\em Phys. Rev. Lett.}
  {\bfseries 38} (1977) 1440--1443}.

\bibitem{Peccei:1977ur}
R.~D. Peccei and H.~R. Quinn, ``{Constraints Imposed by CP Conservation in the
  Presence of Instantons},''
\href{http://dx.doi.org/10.1103/PhysRevD.16.1791}{{\em Phys. Rev.} {\bfseries
  D16} (1977) 1791--1797}.

\bibitem{Weinberg:1977ma}
S.~Weinberg, ``{A New Light Boson?},''
\href{http://dx.doi.org/10.1103/PhysRevLett.40.223}{{\em Phys. Rev. Lett.}
  {\bfseries 40} (1978) 223--226}.

\bibitem{Wilczek:1977pj}
F.~Wilczek, ``{Problem of Strong p and t Invariance in the Presence of
  Instantons},''
\href{http://dx.doi.org/10.1103/PhysRevLett.40.279}{{\em Phys. Rev. Lett.}
  {\bfseries 40} (1978) 279--282}.

\bibitem{Kim:1979if}
J.~E. Kim, ``{Weak Interaction Singlet and Strong CP Invariance},''
\href{http://dx.doi.org/10.1103/PhysRevLett.43.103}{{\em Phys. Rev. Lett.}
  {\bfseries 43} (1979) 103}.

\bibitem{Shifman:1979if}
M.~A. Shifman, A.~I. Vainshtein, and V.~I. Zakharov, ``{Can Confinement Ensure
  Natural CP Invariance of Strong Interactions?},''
\href{http://dx.doi.org/10.1016/0550-3213(80)90209-6}{{\em Nucl. Phys.}
  {\bfseries B166} (1980) 493}.

\bibitem{Zhitnitsky:1980tq}
A.~R. Zhitnitsky, ``{On Possible Suppression of the Axion Hadron Interactions.
  (In Russian)},'' {\em Sov. J. Nucl. Phys.} {\bfseries 31} (1980) 260.
[Yad. Fiz.31,497(1980)].

\bibitem{Dine:1981rt}
M.~Dine, W.~Fischler, and M.~Srednicki, ``{A Simple Solution to the Strong CP
  Problem with a Harmless Axion},''
\href{http://dx.doi.org/10.1016/0370-2693(81)90590-6}{{\em Phys. Lett.}
  {\bfseries B104} (1981) 199--202}.

\bibitem{DiLuzio:2020wdo}
L.~Di~Luzio, M.~Giannotti, E.~Nardi, and L.~Visinelli, ``{The landscape of QCD
  axion models},'' \href{http://dx.doi.org/10.1016/j.physrep.2020.06.002}{{\em
  Phys. Rept.} {\bfseries 870} (2020) 1--117},
  \href{http://arxiv.org/abs/2003.01100}{{\ttfamily arXiv:2003.01100
  [hep-ph]}}.

\bibitem{Irastorza:2018dyq}
I.~G. Irastorza and J.~Redondo, ``{New experimental approaches in the search
  for axion-like particles},''
  \href{http://dx.doi.org/10.1016/j.ppnp.2018.05.003}{{\em Prog. Part. Nucl.
  Phys.} {\bfseries 102} (2018) 89--159},
  \href{http://arxiv.org/abs/1801.08127}{{\ttfamily arXiv:1801.08127
  [hep-ph]}}.

\bibitem{Sikivie:2020zpn}
P.~Sikivie, ``{Invisible Axion Search Methods},''
  \href{http://arxiv.org/abs/2003.02206}{{\ttfamily arXiv:2003.02206
  [hep-ph]}}.

\bibitem{Srednicki:1985xd}
M.~Srednicki, ``{Axion Couplings to Matter. 1. CP Conserving Parts},''
  \href{http://dx.doi.org/10.1016/0550-3213(85)90054-9}{{\em Nucl. Phys. B}
  {\bfseries 260} (1985) 689--700}.

\bibitem{Chang:1993gm}
S.~Chang and K.~Choi, ``{Hadronic axion window and the big bang
  nucleosynthesis},''
  \href{http://dx.doi.org/10.1016/0370-2693(93)90656-3}{{\em Phys. Lett. B}
  {\bfseries 316} (1993) 51--56},
  \href{http://arxiv.org/abs/hep-ph/9306216}{{\ttfamily arXiv:hep-ph/9306216}}.

\bibitem{Choi:2021kuy}
K.~Choi, S.~H. Im, H.~J. Kim, and H.~Seong, ``{Precision axion physics with
  running axion couplings},''
  \href{http://dx.doi.org/10.1007/JHEP08(2021)058}{{\em JHEP} {\bfseries 08}
  (2021) 058}, \href{http://arxiv.org/abs/2106.05816}{{\ttfamily
  arXiv:2106.05816 [hep-ph]}}.

\bibitem{Choi:2017gpf}
K.~Choi, S.~H. Im, C.~B. Park, and S.~Yun, ``{Minimal Flavor Violation with
  Axion-like Particles},''
  \href{http://dx.doi.org/10.1007/JHEP11(2017)070}{{\em JHEP} {\bfseries 11}
  (2017) 070}, \href{http://arxiv.org/abs/1708.00021}{{\ttfamily
  arXiv:1708.00021 [hep-ph]}}.

\bibitem{Chala:2020wvs}
M.~Chala, G.~Guedes, M.~Ramos, and J.~Santiago, ``{Running in the ALPs},''
  \href{http://dx.doi.org/10.1140/epjc/s10052-021-08968-2}{{\em Eur. Phys. J.
  C} {\bfseries 81} no.~2, (2021) 181},
  \href{http://arxiv.org/abs/2012.09017}{{\ttfamily arXiv:2012.09017
  [hep-ph]}}.

\bibitem{Bauer:2020jbp}
M.~Bauer, M.~Neubert, S.~Renner, M.~Schnubel, and A.~Thamm, ``{The Low-Energy
  Effective Theory of Axions and ALPs},''
  \href{http://dx.doi.org/10.1007/JHEP04(2021)063}{{\em JHEP} {\bfseries 04}
  (2021) 063}, \href{http://arxiv.org/abs/2012.12272}{{\ttfamily
  arXiv:2012.12272 [hep-ph]}}.

\bibitem{Bonilla:2021ufe}
J.~Bonilla, I.~Brivio, M.~B. Gavela, and V.~Sanz, ``{One-loop corrections to
  ALP couplings},'' \href{http://dx.doi.org/10.1007/JHEP11(2021)168}{{\em JHEP}
  {\bfseries 11} (2021) 168}, \href{http://arxiv.org/abs/2107.11392}{{\ttfamily
  arXiv:2107.11392 [hep-ph]}}.

\bibitem{DiLuzio:2017ogq}
L.~Di~Luzio, F.~Mescia, E.~Nardi, P.~Panci, and R.~Ziegler, ``{Astrophobic
  Axions},'' \href{http://dx.doi.org/10.1103/PhysRevLett.120.261803}{{\em Phys.
  Rev. Lett.} {\bfseries 120} no.~26, (2018) 261803},
\href{http://arxiv.org/abs/1712.04940}{{\ttfamily arXiv:1712.04940 [hep-ph]}}.

\bibitem{Bjorkeroth:2019jtx}
F.~Bjorkeroth, L.~Di~Luzio, F.~Mescia, E.~Nardi, P.~Panci, and R.~Ziegler,
  ``{Axion-electron decoupling in nucleophobic axion models},''
  \href{http://dx.doi.org/10.1103/PhysRevD.101.035027}{{\em Phys. Rev. D}
  {\bfseries 101} no.~3, (2020) 035027},
  \href{http://arxiv.org/abs/1907.06575}{{\ttfamily arXiv:1907.06575
  [hep-ph]}}.

\bibitem{DiLuzio:2022tyc}
L.~Di~Luzio, F.~Mescia, E.~Nardi, and S.~Okawa, ``{Renormalization group
  effects in astrophobic axion models},''
  \href{http://dx.doi.org/10.1103/PhysRevD.106.055016}{{\em Phys. Rev. D}
  {\bfseries 106} no.~5, (2022) 055016},
  \href{http://arxiv.org/abs/2205.15326}{{\ttfamily arXiv:2205.15326
  [hep-ph]}}.

\bibitem{Branco:2011iw}
G.~C. Branco, P.~M. Ferreira, L.~Lavoura, M.~N. Rebelo, M.~Sher, and J.~P.
  Silva, ``{Theory and phenomenology of two-Higgs-doublet models},''
  \href{http://dx.doi.org/10.1016/j.physrep.2012.02.002}{{\em Phys. Rept.}
  {\bfseries 516} (2012) 1--102},
  \href{http://arxiv.org/abs/1106.0034}{{\ttfamily arXiv:1106.0034 [hep-ph]}}.

\bibitem{Krause:2016oke}
M.~Krause, R.~Lorenz, M.~Muhlleitner, R.~Santos, and H.~Ziesche,
  ``{Gauge-independent Renormalization of the 2-Higgs-Doublet Model},''
  \href{http://dx.doi.org/10.1007/JHEP09(2016)143}{{\em JHEP} {\bfseries 09}
  (2016) 143}, \href{http://arxiv.org/abs/1605.04853}{{\ttfamily
  arXiv:1605.04853 [hep-ph]}}.

\bibitem{ParticleDataGroup:2020ssz}
{\bfseries Particle Data Group} Collaboration, P.~A. Zyla {\em et~al.},
  ``{Review of Particle Physics},''
  \href{http://dx.doi.org/10.1093/ptep/ptaa104}{{\em PTEP} {\bfseries 2020}
  no.~8, (2020) 083C01}.

\bibitem{DiLuzio:2016sbl}
L.~Di~Luzio, F.~Mescia, and E.~Nardi, ``{Redefining the Axion Window},''
  \href{http://dx.doi.org/10.1103/PhysRevLett.118.031801}{{\em Phys. Rev.
  Lett.} {\bfseries 118} no.~3, (2017) 031801},
\href{http://arxiv.org/abs/1610.07593}{{\ttfamily arXiv:1610.07593 [hep-ph]}}.

\bibitem{DiLuzio:2017pfr}
L.~Di~Luzio, F.~Mescia, and E.~Nardi, ``{Window for preferred axion models},''
  \href{http://dx.doi.org/10.1103/PhysRevD.96.075003}{{\em Phys. Rev.}
  {\bfseries D96} no.~7, (2017) 075003},
\href{http://arxiv.org/abs/1705.05370}{{\ttfamily arXiv:1705.05370 [hep-ph]}}.

\bibitem{Bauer:2017ris}
M.~Bauer, M.~Neubert, and A.~Thamm, ``{Collider Probes of Axion-Like
  Particles},'' \href{http://dx.doi.org/10.1007/JHEP12(2017)044}{{\em JHEP}
  {\bfseries 12} (2017) 044}, \href{http://arxiv.org/abs/1708.00443}{{\ttfamily
  arXiv:1708.00443 [hep-ph]}}.

\bibitem{diCortona:2015ldu}
G.~Grilli~di Cortona, E.~Hardy, J.~Pardo~Vega, and G.~Villadoro, ``{The QCD
  axion, precisely},'' \href{http://dx.doi.org/10.1007/JHEP01(2016)034}{{\em
  JHEP} {\bfseries 01} (2016) 034},
\href{http://arxiv.org/abs/1511.02867}{{\ttfamily arXiv:1511.02867 [hep-ph]}}.

\bibitem{Choi:2021ign}
K.~Choi, H.~J. Kim, H.~Seong, and C.~S. Shin, ``{Axion emission from supernova
  with axion-pion-nucleon contact interaction},''
  \href{http://dx.doi.org/10.1007/JHEP02(2022)143}{{\em JHEP} {\bfseries 02}
  (2022) 143}, \href{http://arxiv.org/abs/2110.01972}{{\ttfamily
  arXiv:2110.01972 [hep-ph]}}.

\bibitem{Ho:2022oaw}
S.-Y. Ho, J.~Kim, P.~Ko, and J.-h. Park, ``{Supernova axion emissivity with
  \ensuremath{\Delta}(1232) resonance in heavy baryon chiral perturbation
  theory},'' \href{http://dx.doi.org/10.1103/PhysRevD.107.075002}{{\em Phys.
  Rev. D} {\bfseries 107} no.~7, (2023) 075002},
  \href{http://arxiv.org/abs/2212.01155}{{\ttfamily arXiv:2212.01155
  [hep-ph]}}.

\bibitem{FLAG2023}
{\bfseries Flavour Lattice Averaging Group (FLAG Review 2023 update)}
  Collaboration. {\url{HTTP://FLAG.UNIBE.CH/2021/}}.

\bibitem{Liang:2018pis}
J.~Liang, Y.-B. Yang, T.~Draper, M.~Gong, and K.-F. Liu, ``{Quark spins and
  Anomalous Ward Identity},''
  \href{http://dx.doi.org/10.1103/PhysRevD.98.074505}{{\em Phys. Rev. D}
  {\bfseries 98} no.~7, (2018) 074505},
  \href{http://arxiv.org/abs/1806.08366}{{\ttfamily arXiv:1806.08366
  [hep-ph]}}.

\bibitem{FlavourLatticeAveragingGroupFLAG:2021npn}
{\bfseries Flavour Lattice Averaging Group (FLAG)} Collaboration, Y.~Aoki {\em
  et~al.}, ``{FLAG Review 2021},''
  \href{http://dx.doi.org/10.1140/epjc/s10052-022-10536-1}{{\em Eur. Phys. J.
  C} {\bfseries 82} no.~10, (2022) 869},
  \href{http://arxiv.org/abs/2111.09849}{{\ttfamily arXiv:2111.09849
  [hep-lat]}}.

\bibitem{Gunion:2002zf}
J.~F. Gunion and H.~E. Haber, ``{The CP conserving two Higgs doublet model: The
  Approach to the decoupling limit},''
  \href{http://dx.doi.org/10.1103/PhysRevD.67.075019}{{\em Phys. Rev. D}
  {\bfseries 67} (2003) 075019},
  \href{http://arxiv.org/abs/hep-ph/0207010}{{\ttfamily arXiv:hep-ph/0207010}}.

\bibitem{Bertolini:2014aia}
S.~Bertolini, L.~Di~Luzio, H.~Kole\v{s}ov\'a, and M.~Malinsk\'y, ``{Massive
  neutrinos and invisible axion minimally connected},''
  \href{http://dx.doi.org/10.1103/PhysRevD.91.055014}{{\em Phys. Rev. D}
  {\bfseries 91} no.~5, (2015) 055014},
  \href{http://arxiv.org/abs/1412.7105}{{\ttfamily arXiv:1412.7105 [hep-ph]}}.

\bibitem{Espriu:2015mfa}
D.~Espriu, F.~Mescia, and A.~Renau, ``{Axion-Higgs interplay in the two
  Higgs-doublet model},''
  \href{http://dx.doi.org/10.1103/PhysRevD.92.095013}{{\em Phys. Rev. D}
  {\bfseries 92} no.~9, (2015) 095013},
  \href{http://arxiv.org/abs/1503.02953}{{\ttfamily arXiv:1503.02953
  [hep-ph]}}.

\bibitem{DiLuzio:2016sur}
L.~Di~Luzio, J.~F. Kamenik, and M.~Nardecchia, ``{Implications of perturbative
  unitarity for scalar di-boson resonance searches at LHC},''
  \href{http://dx.doi.org/10.1140/epjc/s10052-017-4594-2}{{\em Eur. Phys. J. C}
  {\bfseries 77} no.~1, (2017) 30},
  \href{http://arxiv.org/abs/1604.05746}{{\ttfamily arXiv:1604.05746
  [hep-ph]}}.

\bibitem{Antusch:2013jca}
S.~Antusch and V.~Maurer, ``{Running quark and lepton parameters at various
  scales},'' \href{http://dx.doi.org/10.1007/JHEP11(2013)115}{{\em JHEP}
  {\bfseries 11} (2013) 115}, \href{http://arxiv.org/abs/1306.6879}{{\ttfamily
  arXiv:1306.6879 [hep-ph]}}.

\bibitem{Carenza:2020cis}
P.~Carenza, B.~Fore, M.~Giannotti, A.~Mirizzi, and S.~Reddy, ``{Enhanced
  Supernova Axion Emission and its Implications},''
  \href{http://dx.doi.org/10.1103/PhysRevLett.126.071102}{{\em Phys. Rev.
  Lett.} {\bfseries 126} no.~7, (2021) 071102},
  \href{http://arxiv.org/abs/2010.02943}{{\ttfamily arXiv:2010.02943
  [hep-ph]}}.

\bibitem{Fischer:2021jfm}
T.~Fischer, P.~Carenza, B.~Fore, M.~Giannotti, A.~Mirizzi, and S.~Reddy,
  ``{Observable signatures of enhanced axion emission from protoneutron
  stars},'' \href{http://dx.doi.org/10.1103/PhysRevD.104.103012}{{\em Phys.
  Rev. D} {\bfseries 104} no.~10, (2021) 103012},
  \href{http://arxiv.org/abs/2108.13726}{{\ttfamily arXiv:2108.13726
  [hep-ph]}}.

\bibitem{Lella:2022uwi}
A.~Lella, P.~Carenza, G.~Lucente, M.~Giannotti, and A.~Mirizzi,
  ``{Proto-neutron stars as cosmic factories for massive
  axion-like-particles},'' \href{http://arxiv.org/abs/2211.13760}{{\ttfamily
  arXiv:2211.13760 [hep-ph]}}.

\bibitem{Raffelt:1982dr}
G.~Raffelt and L.~Stodolsky, ``{New Particles From Nuclear Reactions in the
  Sun},'' \href{http://dx.doi.org/10.1016/0370-2693(82)90680-3}{{\em Phys.
  Lett. B} {\bfseries 119} (1982) 323}.

\bibitem{CAST:2009jdc}
{\bfseries CAST} Collaboration, S.~Andriamonje {\em et~al.}, ``{Search for
  14.4-keV solar axions emitted in the M1-transition of Fe-57 nuclei with
  CAST},'' \href{http://dx.doi.org/10.1088/1475-7516/2009/12/002}{{\em JCAP}
  {\bfseries 12} (2009) 002}, \href{http://arxiv.org/abs/0906.4488}{{\ttfamily
  arXiv:0906.4488 [hep-ex]}}.

\bibitem{Borexino:2012guz}
{\bfseries Borexino} Collaboration, G.~Bellini {\em et~al.}, ``{Search for
  Solar Axions Produced in $p(d,\rm{^3He})A$ Reaction with Borexino
  Detector},'' \href{http://dx.doi.org/10.1103/PhysRevD.85.092003}{{\em Phys.
  Rev. D} {\bfseries 85} (2012) 092003},
  \href{http://arxiv.org/abs/1203.6258}{{\ttfamily arXiv:1203.6258 [hep-ex]}}.

\bibitem{Bhusal:2020bvx}
A.~Bhusal, N.~Houston, and T.~Li, ``{Searching for Solar Axions Using Data from
  the Sudbury Neutrino Observatory},''
  \href{http://dx.doi.org/10.1103/PhysRevLett.126.091601}{{\em Phys. Rev.
  Lett.} {\bfseries 126} no.~9, (2021) 091601},
  \href{http://arxiv.org/abs/2004.02733}{{\ttfamily arXiv:2004.02733
  [hep-ph]}}.

\bibitem{Lucente:2022esm}
G.~Lucente, N.~Nath, F.~Capozzi, M.~Giannotti, and A.~Mirizzi, ``{Probing
  high-energy solar axion flux with a large scintillation neutrino detector},''
  \href{http://dx.doi.org/10.1103/PhysRevD.106.123007}{{\em Phys. Rev. D}
  {\bfseries 106} no.~12, (2022) 123007},
  \href{http://arxiv.org/abs/2209.11780}{{\ttfamily arXiv:2209.11780
  [hep-ph]}}.

\bibitem{DiLuzio:2021qct}
L.~Di~Luzio {\em et~al.}, ``{Probing the axion\textendash{}nucleon coupling
  with the next generation of~axion helioscopes},''
  \href{http://dx.doi.org/10.1140/epjc/s10052-022-10061-1}{{\em Eur. Phys. J.
  C} {\bfseries 82} no.~2, (2022) 120},
  \href{http://arxiv.org/abs/2111.06407}{{\ttfamily arXiv:2111.06407
  [hep-ph]}}.

\bibitem{ParticleDataGroup:2022pth}
{\bfseries Particle Data Group} Collaboration, R.~L. Workman {\em et~al.},
  ``{Review of Particle Physics},''
  \href{http://dx.doi.org/10.1093/ptep/ptac097}{{\em PTEP} {\bfseries 2022}
  (2022) 083C01}.

\bibitem{Hannestad:2005df}
S.~Hannestad, A.~Mirizzi, and G.~Raffelt, ``{New cosmological mass limit on
  thermal relic axions},''
  \href{http://dx.doi.org/10.1088/1475-7516/2005/07/002}{{\em JCAP} {\bfseries
  07} (2005) 002}, \href{http://arxiv.org/abs/hep-ph/0504059}{{\ttfamily
  arXiv:hep-ph/0504059}}.

\bibitem{DiLuzio:2021vjd}
L.~Di~Luzio, G.~Martinelli, and G.~Piazza, ``{Breakdown of chiral perturbation
  theory for the axion hot dark matter bound},''
  \href{http://dx.doi.org/10.1103/PhysRevLett.126.241801}{{\em Phys. Rev.
  Lett.} {\bfseries 126} no.~24, (2021) 241801},
  \href{http://arxiv.org/abs/2101.10330}{{\ttfamily arXiv:2101.10330
  [hep-ph]}}.

\bibitem{Notari:2022zxo}
A.~Notari, F.~Rompineve, and G.~Villadoro, ``{Improved hot dark matter bound on
  the QCD axion},'' \href{http://arxiv.org/abs/2211.03799}{{\ttfamily
  arXiv:2211.03799 [hep-ph]}}.

\bibitem{DiLuzio:2022gsc}
L.~Di~Luzio, J.~Martin~Camalich, G.~Martinelli, J.~A. Oller, and G.~Piazza,
  ``{Axion-pion thermalization rate in unitarized NLO chiral perturbation
  theory},'' \href{http://arxiv.org/abs/2211.05073}{{\ttfamily arXiv:2211.05073
  [hep-ph]}}.

\bibitem{DiLuzio:2021ysg}
L.~Di~Luzio, M.~Fedele, M.~Giannotti, F.~Mescia, and E.~Nardi, ``{Stellar
  evolution confronts axion models},''
  \href{http://dx.doi.org/10.1088/1475-7516/2022/02/035}{{\em JCAP} {\bfseries
  02} (2022) 035}, \href{http://arxiv.org/abs/2109.10368}{{\ttfamily
  arXiv:2109.10368 [hep-ph]}}.

\bibitem{Arvanitaki:2014dfa}
A.~Arvanitaki and A.~A. Geraci, ``{Resonantly Detecting Axion-Mediated Forces
  with Nuclear Magnetic Resonance},''
  \href{http://dx.doi.org/10.1103/PhysRevLett.113.161801}{{\em Phys. Rev.
  Lett.} {\bfseries 113} no.~16, (2014) 161801},
\href{http://arxiv.org/abs/1403.1290}{{\ttfamily arXiv:1403.1290 [hep-ph]}}.

\bibitem{Garcon:2017ixh}
A.~Garcon {\em et~al.}, ``{The Cosmic Axion Spin Precession Experiment
  (CASPEr): a dark-matter search with nuclear magnetic resonance},''
  \href{http://arxiv.org/abs/1707.05312}{{\ttfamily arXiv:1707.05312
  [physics.ins-det]}}.

\bibitem{LUX:2017glr}
{\bfseries LUX} Collaboration, D.~S. Akerib {\em et~al.}, ``{First Searches for
  Axions and Axionlike Particles with the LUX Experiment},''
  \href{http://dx.doi.org/10.1103/PhysRevLett.118.261301}{{\em Phys. Rev.
  Lett.} {\bfseries 118} no.~26, (2017) 261301},
  \href{http://arxiv.org/abs/1704.02297}{{\ttfamily arXiv:1704.02297
  [astro-ph.CO]}}.

\bibitem{PandaX-II:2020udv}
{\bfseries PandaX-II} Collaboration, X.~Zhou {\em et~al.}, ``{A Search for
  Solar Axions and Anomalous Neutrino Magnetic Moment with the Complete
  PandaX-II Data},''
  \href{http://dx.doi.org/10.1088/0256-307X/38/10/109902}{{\em Chin. Phys.
  Lett.} {\bfseries 38} no.~1, (2021) 011301},
  \href{http://arxiv.org/abs/2008.06485}{{\ttfamily arXiv:2008.06485
  [hep-ex]}}. [Erratum: Chin.Phys.Lett. 38, 109902 (2021)].

\bibitem{XENON:2022ltv}
{\bfseries XENON} Collaboration, E.~Aprile {\em et~al.}, ``{Search for New
  Physics in Electronic Recoil Data from XENONnT},''
  \href{http://dx.doi.org/10.1103/PhysRevLett.129.161805}{{\em Phys. Rev.
  Lett.} {\bfseries 129} no.~16, (2022) 161805},
  \href{http://arxiv.org/abs/2207.11330}{{\ttfamily arXiv:2207.11330
  [hep-ex]}}.

\bibitem{Sikivie:1983ip}
P.~Sikivie, ``{Experimental Tests of the Invisible Axion},''
  \href{http://dx.doi.org/10.1103/PhysRevLett.51.1415}{{\em Phys. Rev. Lett.}
  {\bfseries 51} (1983) 1415--1417}. [Erratum: Phys.Rev.Lett. 52, 695 (1984)].

\bibitem{Ayala:2014pea}
A.~Ayala, I.~Dominguez, M.~Giannotti, A.~Mirizzi, and O.~Straniero,
  ``{Revisiting the bound on axion-photon coupling from Globular Clusters},''
  \href{http://dx.doi.org/10.1103/PhysRevLett.113.191302}{{\em Phys. Rev.
  Lett.} {\bfseries 113} no.~19, (2014) 191302},
  \href{http://arxiv.org/abs/1406.6053}{{\ttfamily arXiv:1406.6053
  [astro-ph.SR]}}.

\bibitem{Straniero:2015nvc}
O.~Straniero, A.~Ayala, M.~Giannotti, A.~Mirizzi, and I.~Dominguez,
  \href{http://dx.doi.org/10.3204/DESY-PROC-2015-02/straniero\_oscar}{``{Axion-Photon
  Coupling: Astrophysical Constraints},''} in {\em {11th Patras Workshop on
  Axions, WIMPs and WISPs}}, pp.~77--81.
\newblock 2015.

\bibitem{Giannotti:2015kwo}
M.~Giannotti, I.~Irastorza, J.~Redondo, and A.~Ringwald, ``{Cool WISPs for
  stellar cooling excesses},''
  \href{http://dx.doi.org/10.1088/1475-7516/2016/05/057}{{\em JCAP} {\bfseries
  05} (2016) 057}, \href{http://arxiv.org/abs/1512.08108}{{\ttfamily
  arXiv:1512.08108 [astro-ph.HE]}}.

\bibitem{Giannotti:2017hny}
M.~Giannotti, I.~G. Irastorza, J.~Redondo, A.~Ringwald, and K.~Saikawa,
  ``{Stellar Recipes for Axion Hunters},''
  \href{http://dx.doi.org/10.1088/1475-7516/2017/10/010}{{\em JCAP} {\bfseries
  1710} no.~10, (2017) 010},
\href{http://arxiv.org/abs/1708.02111}{{\ttfamily arXiv:1708.02111 [hep-ph]}}.

\bibitem{Straniero:2020iyi}
O.~Straniero, C.~Pallanca, E.~Dalessandro, I.~Dominguez, F.~Ferraro,
  M.~Giannotti, A.~Mirizzi, and L.~Piersanti, ``{The RGB tip of galactic
  globular clusters and the revision of the bound of the axion-electron
  coupling},'' \href{http://arxiv.org/abs/2010.03833}{{\ttfamily
  arXiv:2010.03833 [astro-ph.SR]}}.

\bibitem{Buschmann:2021juv}
M.~Buschmann, C.~Dessert, J.~W. Foster, A.~J. Long, and B.~R. Safdi, ``{Upper
  Limit on the QCD Axion Mass from Isolated Neutron Star Cooling},''
  \href{http://dx.doi.org/10.1103/PhysRevLett.128.091102}{{\em Phys. Rev.
  Lett.} {\bfseries 128} no.~9, (2022) 091102},
  \href{http://arxiv.org/abs/2111.09892}{{\ttfamily arXiv:2111.09892
  [hep-ph]}}.

\bibitem{Carenza:2019pxu}
P.~Carenza, T.~Fischer, M.~Giannotti, G.~Guo, G.~Mart\'\i{}nez-Pinedo, and
  A.~Mirizzi, ``{Improved axion emissivity from a supernova via nucleon-nucleon
  bremsstrahlung},''
  \href{http://dx.doi.org/10.1088/1475-7516/2019/10/016}{{\em JCAP} {\bfseries
  10} no.~10, (2019) 016}, \href{http://arxiv.org/abs/1906.11844}{{\ttfamily
  arXiv:1906.11844 [hep-ph]}}. [Erratum: JCAP 05, E01 (2020)].

\bibitem{Kolb:1990vq}
E.~W. Kolb and M.~S. Turner,
  \href{http://dx.doi.org/10.1201/9780429492860}{{\em {The Early Universe}}},
  vol.~69.
\newblock 1990.

\bibitem{Cyburt:2015mya}
R.~H. Cyburt, B.~D. Fields, K.~A. Olive, and T.-H. Yeh, ``{Big Bang
  Nucleosynthesis: 2015},''
  \href{http://dx.doi.org/10.1103/RevModPhys.88.015004}{{\em Rev. Mod. Phys.}
  {\bfseries 88} (2016) 015004},
  \href{http://arxiv.org/abs/1505.01076}{{\ttfamily arXiv:1505.01076
  [astro-ph.CO]}}.

\bibitem{Planck:2018nkj}
{\bfseries Planck} Collaboration, N.~Aghanim {\em et~al.}, ``{Planck 2018
  results. I. Overview and the cosmological legacy of Planck},''
  \href{http://dx.doi.org/10.1051/0004-6361/201833880}{{\em Astron. Astrophys.}
  {\bfseries 641} (2020) A1}, \href{http://arxiv.org/abs/1807.06205}{{\ttfamily
  arXiv:1807.06205 [astro-ph.CO]}}.

\bibitem{Planck:2018vyg}
{\bfseries Planck} Collaboration, N.~Aghanim {\em et~al.}, ``{Planck 2018
  results. VI. Cosmological parameters},''
  \href{http://dx.doi.org/10.1051/0004-6361/201833910}{{\em Astron. Astrophys.}
  {\bfseries 641} (2020) A6}, \href{http://arxiv.org/abs/1807.06209}{{\ttfamily
  arXiv:1807.06209 [astro-ph.CO]}}. [Erratum: Astron.Astrophys. 652, C4
  (2021)].

\bibitem{CMB-S4:2016ple}
{\bfseries CMB-S4} Collaboration, K.~N. Abazajian {\em et~al.}, ``{CMB-S4
  Science Book, First Edition},''
  \href{http://arxiv.org/abs/1610.02743}{{\ttfamily arXiv:1610.02743
  [astro-ph.CO]}}.

\bibitem{SimonsObservatory:2018koc}
{\bfseries Simons Observatory} Collaboration, P.~Ade {\em et~al.}, ``{The
  Simons Observatory: Science goals and forecasts},''
  \href{http://dx.doi.org/10.1088/1475-7516/2019/02/056}{{\em JCAP} {\bfseries
  02} (2019) 056}, \href{http://arxiv.org/abs/1808.07445}{{\ttfamily
  arXiv:1808.07445 [astro-ph.CO]}}.

\bibitem{Caloni:2022uya}
L.~Caloni, M.~Gerbino, M.~Lattanzi, and L.~Visinelli, ``{Novel cosmological
  bounds on thermally-produced axion-like particles},''
  \href{http://dx.doi.org/10.1088/1475-7516/2022/09/021}{{\em JCAP} {\bfseries
  09} (2022) 021}, \href{http://arxiv.org/abs/2205.01637}{{\ttfamily
  arXiv:2205.01637 [astro-ph.CO]}}.

\bibitem{DEramo:2022nvb}
F.~D'Eramo, E.~Di~Valentino, W.~Giar\`e, F.~Hajkarim, A.~Melchiorri, O.~Mena,
  F.~Renzi, and S.~Yun, ``{Cosmological bound on the QCD axion mass, redux},''
  \href{http://dx.doi.org/10.1088/1475-7516/2022/09/022}{{\em JCAP} {\bfseries
  09} (2022) 022}, \href{http://arxiv.org/abs/2205.07849}{{\ttfamily
  arXiv:2205.07849 [astro-ph.CO]}}.

\bibitem{Ferreira:2020bpb}
R.~Z. Ferreira, A.~Notari, and F.~Rompineve,
  ``{Dine-Fischler-Srednicki-Zhitnitsky axion in the CMB},''
  \href{http://dx.doi.org/10.1103/PhysRevD.103.063524}{{\em Phys. Rev. D}
  {\bfseries 103} no.~6, (2021) 063524},
  \href{http://arxiv.org/abs/2012.06566}{{\ttfamily arXiv:2012.06566
  [hep-ph]}}.

\bibitem{IAXO:2020wwp}
{\bfseries IAXO} Collaboration, A.~Abeln {\em et~al.}, ``{Conceptual design of
  BabyIAXO, the intermediate stage towards the International Axion
  Observatory},'' \href{http://dx.doi.org/10.1007/JHEP05(2021)137}{{\em JHEP}
  {\bfseries 05} (2021) 137}, \href{http://arxiv.org/abs/2010.12076}{{\ttfamily
  arXiv:2010.12076 [physics.ins-det]}}.

\bibitem{IAXO:2019mpb}
{\bfseries IAXO} Collaboration, E.~Armengaud {\em et~al.}, ``{Physics potential
  of the International Axion Observatory (IAXO)},''
  \href{http://dx.doi.org/10.1088/1475-7516/2019/06/047}{{\em JCAP} {\bfseries
  06} (2019) 047}, \href{http://arxiv.org/abs/1904.09155}{{\ttfamily
  arXiv:1904.09155 [hep-ph]}}.

\bibitem{Georgi:1986df}
H.~Georgi, D.~B. Kaplan, and L.~Randall, ``{Manifesting the Invisible Axion at
  Low-energies},'' \href{http://dx.doi.org/10.1016/0370-2693(86)90688-X}{{\em
  Phys. Lett. B} {\bfseries 169} (1986) 73--78}.

\bibitem{Chetyrkin:1998mw}
K.~G. Chetyrkin, B.~A. Kniehl, M.~Steinhauser, and W.~A. Bardeen, ``{Effective
  QCD interactions of CP odd Higgs bosons at three loops},''
  \href{http://dx.doi.org/10.1016/S0550-3213(98)00594-X}{{\em Nucl. Phys. B}
  {\bfseries 535} (1998) 3--18},
  \href{http://arxiv.org/abs/hep-ph/9807241}{{\ttfamily arXiv:hep-ph/9807241}}.

\bibitem{Choi:2020rgn}
K.~Choi, S.~H. Im, and C.~S. Shin, ``{Recent progress in physics of axions or
  axion-like particles},'' \href{http://arxiv.org/abs/2012.05029}{{\ttfamily
  arXiv:2012.05029 [hep-ph]}}.

\end{thebibliography}\endgroup

\end{document}